\newcommand{\ThesisAuthorFirstName}{Jannik}
\newcommand{\ThesisAuthorFamilyName}{Nitschke}
\newcommand{\ThesisAuthorImmatriculationNumber}{1636338}
\newcommand{\ThesisAuthorUniversity}{University of Siegen}
\newcommand{\ThesisAuthorFaculty}{Intelligent Systems Group}
\newcommand{\ThesisAuthorDegree}{Bachelor of Science}
\newcommand{\ThesisAuthorStudies}{Computer Science}
\newcommand{\ThesisAuthorStudiesSpecialization}{Complex and Intelligent
Software Systems}
\newcommand{\ThesisAuthorRegulationVersion}{2021}

\newcommand{\ThesisTitle}{The Environmental Impact of Ensemble Techniques in Recommender Systems}
\newcommand{\ThesisType}{Bachelor Thesis}
\newcommand{\ThesisStartDate}{01.07.2025}
\newcommand{\ThesisEndDate}{04.11.2025}

\newcommand{\ThesisExaminerATitle}{Prof.\,Dr.} 
\newcommand{\ThesisExaminerAFirstName}{Joeran}
\newcommand{\ThesisExaminerAFamilyName}{Beel}
\newcommand{\ThesisExaminerADepartment}{Intelligent Systems Group}

\newcommand{\ThesisExaminerBTitle}{B.\,Sc.} 
\newcommand{\ThesisExaminerBFirstName}{Moritz}
\newcommand{\ThesisExaminerBFamilyName}{Baumgart}
\newcommand{\ThesisExaminerBDepartment}{Intelligent Systems Group}

\documentclass[12pt]{book}

\usepackage[T1]{fontenc}
\usepackage[utf8]{inputenc}
\usepackage[table]{xcolor}
\usepackage[paper=a4paper, tmargin=3cm, bmargin=3cm, lmargin=2.5cm, rmargin=2.5cm, bindingoffset=1cm]{geometry}
\usepackage{tabularx}
\usepackage{graphicx}
\usepackage[super]{nth}
\usepackage[numbers,sort&compress]{natbib}
\usepackage{float}
\usepackage{xurl}
\usepackage{hyperref}
\usepackage{calligra}

\begin{document}

\pagestyle{empty}

{\centering
\par\vspace*{1cm}
{\scshape\large \ThesisAuthorFaculty{}\par\vspace{-0.15cm}}
{\scshape\LARGE \ThesisAuthorUniversity{}\par}
\vspace{1.0cm}
{\huge\bfseries \ThesisTitle{}\par}
\vspace{1.0cm}
{\scshape\LARGE \ThesisType{}\par}
{\itshape\large \ThesisAuthorStudies{}\par\vspace*{-0.07cm}}
{\itshape\large \ThesisAuthorStudiesSpecialization{}\par\vspace*{-0.07cm}}
{\itshape\LARGE \ThesisAuthorFirstName{} \ThesisAuthorFamilyName{}\par}
{\itshape\large \ThesisAuthorImmatriculationNumber{}\par}
\vspace{1.0cm}
{\ThesisEndDate{}}
\vfill

\includegraphics[width=0.5\textwidth]{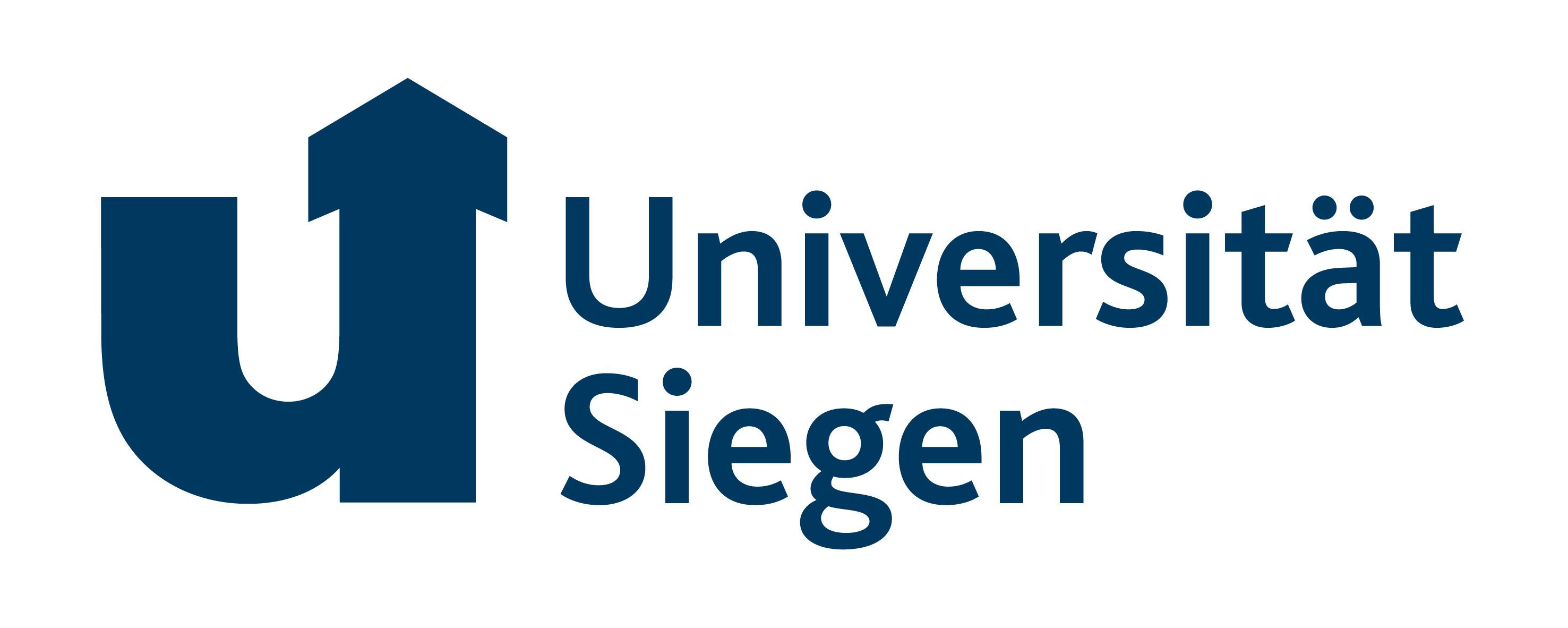}\par\vspace{0.1\textwidth}
\begin{figure}[h]
    \centering
    \includegraphics[width=0.25\textwidth]{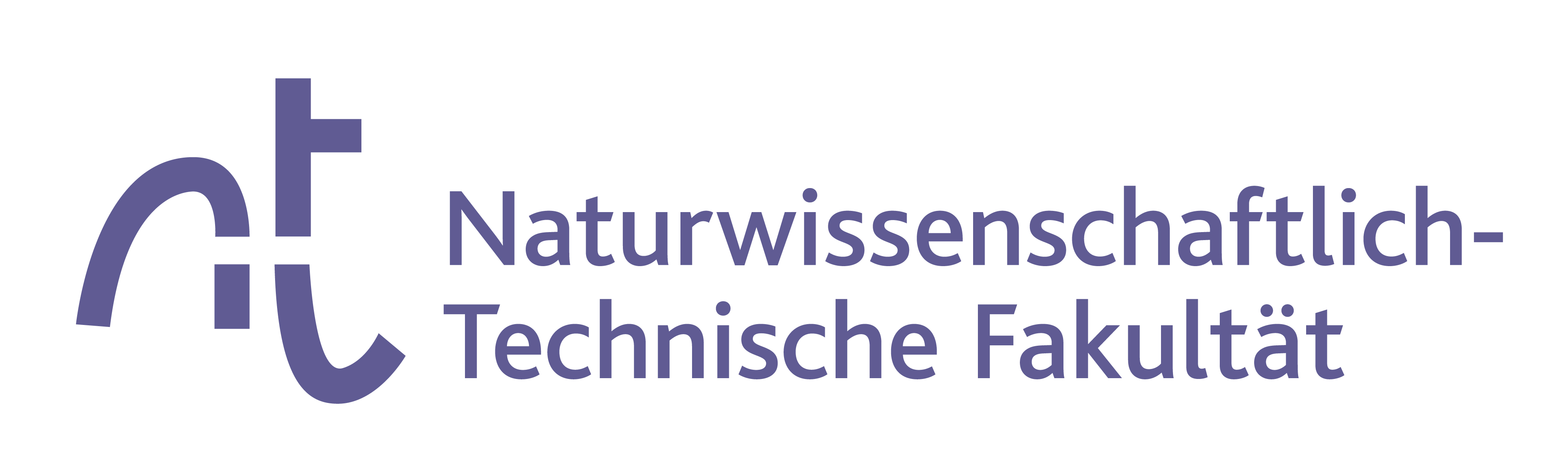}
    \includegraphics[width=0.1\textwidth]{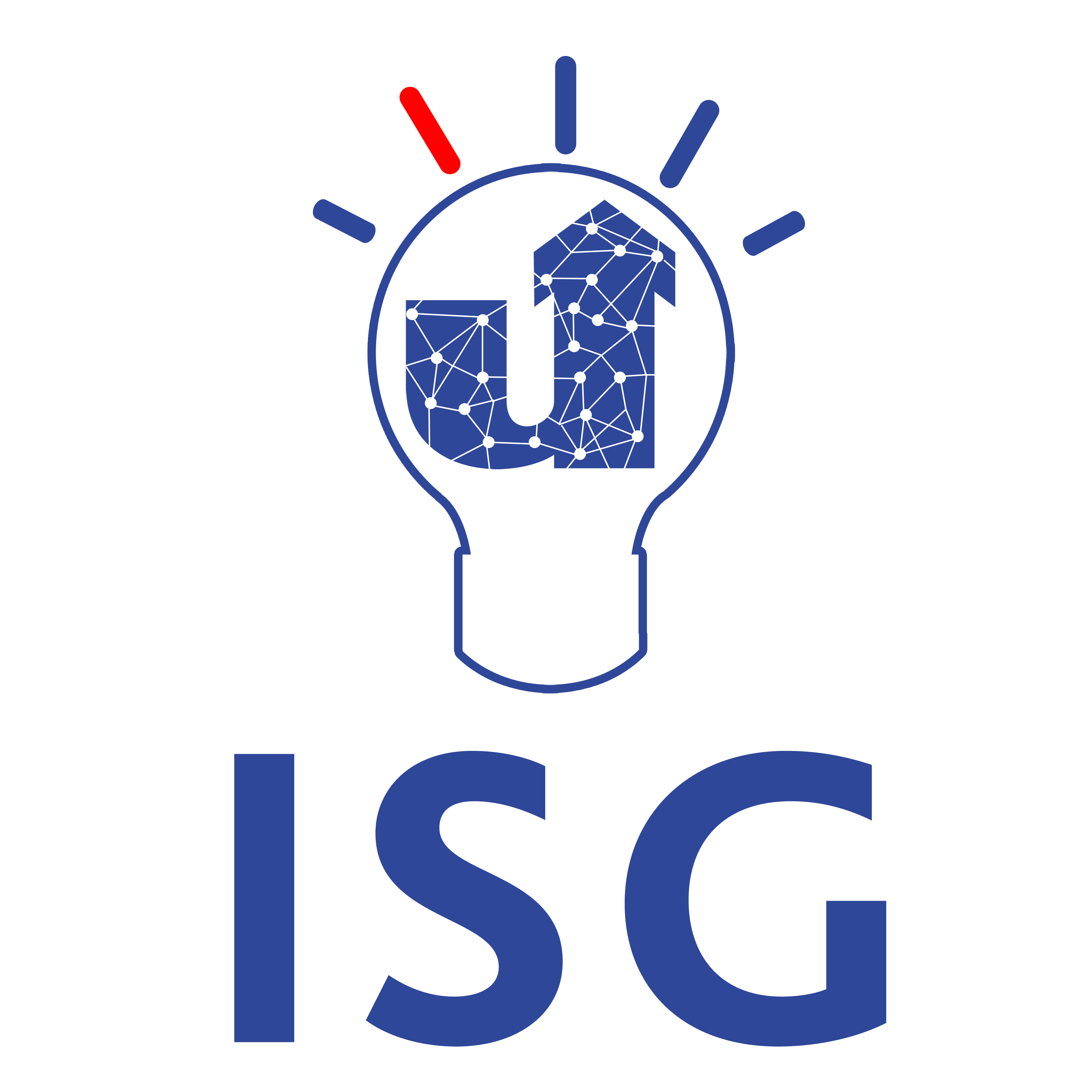}
\end{figure}

\begin{tabularx}{\textwidth}{@{} l X r @{}}
	{\small \textbf{Examiners}} & &  \\
	{\ThesisExaminerATitle{} \ThesisExaminerAFirstName{} \textsc{\ThesisExaminerAFamilyName{}}} & & {\ThesisExaminerADepartment} \\
	{\ThesisExaminerBTitle{} \ThesisExaminerBFirstName{} \textsc{\ThesisExaminerBFamilyName{}}} & & {\ThesisExaminerBDepartment} \\
\end{tabularx}\par}

\clearpage{}

\noindent This \ThesisType{} is handed in according to the requirements of the \ThesisAuthorUniversity{} for the study program \ThesisAuthorDegree{} \ThesisAuthorStudies{} of the year \ThesisAuthorRegulationVersion{} (PO\,\ThesisAuthorRegulationVersion{}).\par

\vfill

\begin{tabularx}{\textwidth}{@{} l X @{}}
	{\small \textbf{Process period}} & \\
	{\ThesisStartDate{} to \ThesisEndDate{}} & \\[0.4cm]
	{\small \textbf{Examiners}} & \\
	{\ThesisExaminerATitle{} \ThesisExaminerAFirstName{} \textsc{\ThesisExaminerAFamilyName{}}} & \\
	{\ThesisExaminerBTitle{} \ThesisExaminerBFirstName{} \textsc{\ThesisExaminerBFamilyName{}}} & \\
\end{tabularx}\par

\clearpage{}

\section*{\underline{Statutory Declaration}}

\vspace{2\baselineskip}
I affirm that I have written my thesis (in the case of a group thesis, my appropriately marked part of the thesis) independently and that I have not used any sources or aids other than those indicated, and that I have clearly indicated citations.

\vspace{\baselineskip}
\noindent
All passages that are taken from other works in terms of wording or meaning (including translations) have been clearly marked as borrowed in each individual case, with precise indication of the source (including the World Wide Web and other electronic data collections).
This also applies to attached drawings, pictorial representations, sketches and the like.
I take note that the proven omission of the indication of origin will be considered as attempted deception.

\vspace{3\baselineskip}

\noindent
\begin{tabbing}
\makebox[0.45\linewidth][l]{Siegen, 04.11.2025} \hspace{0.1\linewidth} \= \makebox[0.45\linewidth][l]{{\fontsize{20}{24}\selectfont\calligra Jannik Nitschke}}
\\
\rule{0.45\linewidth}{0.5pt} \> \rule{0.45\linewidth}{0.5pt}
\\
Location, Date \> Signature
\end{tabbing}

\chapter*{Abstract}

Ensemble techniques in recommender systems have demonstrated accuracy improvements of 10-30\%, yet their environmental impact remains unmeasured. While deep learning recommendation algorithms can generate up to 3,297 kg CO\textsubscript{2} per paper, ensemble methods have not been sufficiently evaluated for energy consumption. This thesis investigates how ensemble techniques influence environmental impact compared to single optimized models.

We conducted 93 experiments across two frameworks (Surprise for rating prediction, LensKit for ranking) on four datasets spanning 100,000 to 7.8 million interactions. We evaluated four ensemble strategies (Average, Weighted, Stacking/Rank Fusion, Top Performers) against simple baselines and optimized single models, measuring energy consumption with a smart plug.

Results revealed a non-linear accuracy-energy relationship. Ensemble methods achieved 0.3-5.7\% accuracy improvements while consuming 19-2,549\% more energy depending on dataset size and strategy. The Top Performers ensemble showed best efficiency: 0.96\% RMSE improvement with 18.8\% energy overhead on MovieLens-1M, and 5.7\% NDCG improvement with 103\% overhead on MovieLens-100K. Exhaustive averaging strategies consumed 88-270\% more energy for comparable gains. On the largest dataset (Anime, 7.8M interactions), the Surprise ensemble consumed 2,005\% more energy (0.21 Wh vs. 0.01 Wh) for 1.2\% accuracy improvement, producing 53.8 mg CO\textsubscript{2} versus 2.6 mg CO\textsubscript{2} for the single model.

This research provides one of the first systematic measurements of energy and carbon footprint for ensemble recommender systems, demonstrates that selective strategies offer superior efficiency over exhaustive averaging, and identifies scalability limitations at industrial scale. These findings enable informed decisions about sustainable algorithm selection in recommender systems.
\\
\\
\noindent \textit{Keywords: Recommender Systems, Ensemble Techniques, Environmental Impact, Energy Consumption, Carbon Footprint, Machine Learning, Sustainability, Green AI} 

\renewcommand{\contentsname}{Table of Contents}
\tableofcontents

\pagestyle{plain}

\chapter{Introduction}
\section{Background}
Over 99\% of climate scientists agree that global warming is real and mainly caused by humans~\citep{climate_consensus_lynas2021}. Machine learning itself contributes significantly to this problem, with a rapidly growing carbon footprint~\citep{green_ai_schwartz2019, computing_carbon_gupta2021}. The computing infrastructure required for training and deploying machine learning models consumes high amounts of electricity. As the field continues to expand, with increasingly large models and datasets, the environmental impact grows proportionally~\citep{green_ai_schwartz2019}. Given the urgency of climate action, one would expect researchers to report the environmental cost of their algorithms alongside accuracy metrics. Yet this remains far from standard practice, particularly in the field of recommender systems.
\\
\\
\noindent Recommender systems have become ubiquitous across digital platforms, from e-commerce to streaming services~\citep{recsys_comprehensive_review_raza2025}. They impact our everyday life. These systems analyze user behavior and preferences to recommend relevant items. By doing so, they significantly influence user experience or business outcomes. As competition grows, researchers continuously seek methods to improve recommendation accuracy. Ensemble techniques have achieved promising results by combining predictions from multiple models to achieve superior accuracy compared to single models~\citep{ensemble_greedy_mehta2024, ensemble_graph_embedding_forouzandeh2020}. The underlying strategy is straightforward: all the different models capture different patterns in user behavior, and their combination can leverage these strengths.
\\
\\
\noindent Recent research shows how effective ensemble methods in recommender systems are. \citet{ensemble_greedy_mehta2024} achieved a 21.7\% accuracy improvement across five datasets through greedy ensemble selection, measuring NDCG, precision, and recall across various configurations. Similarly, \citet{ensemble_graph_embedding_forouzandeh2020} report consistent accuracy gains of 10-30\% in movie recommendations by combining collaborative filtering with graph embedding techniques. These improvements are substantial and represent meaningful advances in recommendation accuracy. However, both of these studies focus only on accuracy metrics. Energy consumption is not measured. Carbon footprint is not reported and computational cost stays unquantified.
\\
\\
\noindent Leaving out all this data is problematic given the environmental cost of machine learning. \citet{recsys_carbon_footprint_vente2024} did systematic measurement of carbon footprints across 79 recommendation algorithms. Their findings reveal that deep learning approaches generate 3,297 kg CO\textsubscript{2} equivalents per research paper. This number exceeds emissions from a transatlantic flight between New York and Melbourne. These measurements highlight significant variation in environmental impact across different algorithmic approaches. However, ensemble methods remain excluded from this analysis. To our knowledge, no study systematically examines whether combining multiple models proportionally multiplies the environmental cost, or whether the overhead is negligible compared to training individual base models.
\\
\\
\noindent The question becomes particularly urgent as ensemble methods gain popularity~\citep{recsys_comprehensive_review_raza2025}. On one hand, if ensembles multiply energy consumption by the number of base models, a ten-model ensemble would consume ten times the energy of a single model. This would create a whole new view of the trade-offs involved when using recommender systems. On the other hand, if the ensemble overhead is minimal, for instance, if model combinations share computational resources, then ensemble techniques might offer an acceptable trade-off. Without sufficient empirical measurements, researchers cannot make informed decisions about deploying ensemble techniques in production systems or their research.

\section{Research Problem}

The problem is quantifying the accuracy-energy trade-off for ensemble methods in recommender systems. If using ensembles improves NDCG by 20\%, does it consume 2× the energy of a single optimized model, 10×, or 100×? The answer has many practical implications in different scenarios. Consider a streaming service deploying recommendations to millions of users daily. A 15\% accuracy improvement might increase user engagement and revenue. But if this improvement requires 50× more energy, the environmental and financial costs may outweigh the benefits. Conversely, if the same accuracy gain requires only 2× more energy, the trade-off becomes much more favorable. Or what's the preferred approach for an environmentally conscious startup aiming to minimize its carbon footprint? Understanding how ensemble techniques impact energy consumption relative to accuracy gains is crucial for making sustainable model and deployment choices.
\\ 
\\
\noindent Currently, practitioners in these fields face these decisions without data. Should they implement weighted ensembles, which assign different importance to each base model? Should they use stacking approaches, which train a meta-learner to combine predictions? Should they select only top-performing models for the ensemble? Each strategy offers different accuracy-energy profiles and trade-offs, but to our knowledge, no systematic comparison exists~\citep{ensemble_model_selection_nijkamp2025, ensemble_accuracy_energy_omar2024}. Furthermore, the energy overhead could vary across different contexts. Large datasets might benefit more from ensemble techniques, which could justify higher computational costs. Sparse datasets might show no significant improvement. Explicit rating prediction tasks differ from implicit feedback ranking. Different evaluation metrics, e.g., RMSE versus NDCG, reflecting different recommendation scenarios with distinct accuracy-energy trade-offs. Understanding these varied factors is important for sustainable algorithm selection in real-world scenarios.

\section{Research Question}

Therefore, this thesis addresses the identified gap by answering the following research question:
\\ 
\\
\noindent
\textit{How do ensemble techniques in recommender systems influence the environmental impact compared to a single optimized model?}

\section{Research Objective and Approach}

Our research objective is to systematically evaluate the environmental impact of ensemble techniques in recommender systems and quantify the accuracy-energy trade-off compared to single models. We conduct controlled experiments measuring both recommendation accuracy and energy consumption across multiple configurations, enabling direct comparison between ensemble methods and single optimized models. 
\\ 
\\
\noindent
We use two established recommender systems frameworks representing different recommendation pipelines. The Surprise library~\citep{surprise_joss_hug2020} evaluates explicit rating prediction, where users provide numerical ratings for items, measuring performance using Root Mean Square Error (RMSE). The LensKit framework~\citep{lenskit_documentation} evaluates implicit feedback ranking, where user preferences are inferred from behavioral signals like clicks or purchases, measuring performance using Normalized Discounted Cumulative Gain (NDCG)~\citep{ndcg_metric_jarvelin2002}.
\\ 
\\
\noindent
Our experiments span four benchmark datasets with varying characteristics. 
Movie\-Lens-\allowbreak 100K and Movie\-Lens-\allowbreak 1M contain 100,000 and one million ratings respectively, representing different scales of explicit feedback~\citep{movielens_dataset_harper2015}. ModCloth provides fashion e-commerce data with implicit feedback, while Anime contains sparse ratings for animes~\citep{recbole_datasets}. We implement four ensemble strategies: Average ensembles, Weighted ensembles, Stacking ensembles, and Top Performers ensembles. Each strategy is compared against a baseline and single optimized model.
\\ 
\\
\noindent
Energy measurement uses the EMERS tool with a Shelly Plug S smart plug to capture total system power consumption~\citep{emers_tool_wegmeth2024, power_meters_comparison_jay2023}. For each configuration, we report accuracy (RMSE or NDCG), energy consumption (watt-hours), and carbon footprint (kg CO\textsubscript{2} equivalent). This comprehensive reporting enables us to answer critical questions: How much additional energy does a 10\% accuracy improvement require? Which ensemble strategies offer the best accuracy per watt-hour? Do these relationships hold across different datasets and recommendation tasks? These insights ultimately guide sustainable algorithm selection in recommender systems.
\chapter{Background}

This chapter provides foundational background information on key concepts that are essential for understanding this thesis but are generally well-established in the literature. Readers familiar with recommender systems, ensemble learning, and machine learning may skip this section.

\section{Recommender Systems}
\label{sec:RecommenderSystems}

Recommender systems are information filtering systems that predict user preferences and suggest relevant items from large catalogs~\cite{recsys_comprehensive_review_raza2025}. These systems are widely deployed across e-commerce, streaming platforms, and social media to address information overload and enhance user experience. The two primary recommendation tasks are:

\begin{itemize}
    \item \textbf{Rating Prediction:} Predicting explicit ratings (e.g., 1-5 stars) that users would assign to items
    \item \textbf{Top-N Recommendation:} Generating ranked lists of N items most likely to be relevant to each user
\end{itemize}

\noindent Common algorithmic approaches include collaborative filtering (user-based and item-based), content-based filtering, matrix factorization, and neural network-based methods. For an in-depth comprehensive overview of recommender system architectures and algorithms, see~\cite{recsys_comprehensive_review_raza2025}.

\section{Ensemble Techniques}
\label{sec:EnsembleTechniques}

Ensemble techniques combine predictions from multiple models to achieve better performance than individual models~\cite{ensemble_greedy_mehta2024}. In recommender systems, ensembles can combine diverse algorithms to improve recommendation quality and robustness. This study employs four ensemble strategies:

\begin{itemize}
    \item \textbf{Average:} Combining predictions through unweighted mean of all base model outputs
    \item \textbf{Weighted:} Combining predictions with learnable weights assigned to each base model
    \item \textbf{Stacking:} Training a meta-learner on base model predictions to learn optimal combination
    \item \textbf{Top Performers:} Averaging predictions from only the best-performing individual models
\end{itemize}

\noindent Recent work has explored greedy selection methods for ensemble construction~\cite{ensemble_greedy_mehta2024} and graph embedding approaches~\cite{ensemble_graph_embedding_forouzandeh2020}.

\section{Data Splitting Strategies}
\label{sec:DataSplittingStrategies}

Proper data splitting is crucial for a reliable evaluation of recommender systems. The choice of splitting strategy depends on the evaluation task and can significantly impact the final accuracy results~\cite{random_seeds_wegmeth2024}. This study employs two splitting strategies:
\\
\\
\noindent \textbf{Global Random Split} randomly partitions all user-item interactions into training and testing sets, typically with an 80/20 ratio. Each interaction has an equal probability of being assigned to either set, regardless of which user it belongs to. This approach is commonly used for rating prediction tasks where the goal is to predict explicit ratings. However, it does not guarantee that all users appear in both sets, which can be problematic for ranking evaluation.
\\
\\
\noindent\textbf{Per-User Random Split} ensures that each user appears in both training and testing sets by randomly splitting each user's interactions individually. For each user, a part of their items (e.g., 20\%) is randomly selected for the test set, while the remaining items form the training set. This strategy is essential for ranking-based evaluation, as it ensures that recommendations can be generated and evaluated for all users. Additionally, minimum thresholds can be applied e.g., requiring users to have at least 10 interactions to filter out users with insufficient data~\cite{random_seeds_wegmeth2024}.
\\
\\
\noindent Both strategies use fixed random seeds to ensure reproducibility across experiments~\cite{random_seeds_wegmeth2024}. The choice between global and per-user splitting significantly affects evaluation reliability, particularly when comparing different algorithms or ensemble configurations.

\section{Cross-Validation}
\label{sec:CrossValidation}

Cross-validation is a resampling technique used to obtain the generalization performance of machine learning models. In k-fold cross-validation, the dataset is partitioned into k equal-sized folds. The model is trained on k-1 folds and validated on the remaining fold. In a scenario with 5-fold cross-validation, the process is repeated 5 times, with each fold serving as the validation set once. The final performance metric is obtained by averaging the results across all k iterations~\cite{crossfold_bengio_2004}.

\section{Evaluation Metrics}
\label{sec:EvaluationMetrics}

Recommender systems are evaluated using different metrics depending on the task. For rating prediction, \textbf{Root Mean Squared Error (RMSE)} measures prediction accuracy:
\begin{equation}
RMSE = \sqrt{\frac{1}{n}\sum_{i=1}^{n}(y_i - \hat{y}_i)^2}
\end{equation}
where $n$ is the number of predictions, $y_i$ is the actual rating, and $\hat{y}_i$ is the predicted rating.
\\
\\
\noindent For ranking-based recommendation, this study uses three complementary metrics:

\noindent \textbf{Normalized Discounted Cumulative Gain (NDCG)}~\cite{ndcg_metric_jarvelin2002} measures ranking quality with position-based discounting:
\begin{equation}
NDCG@k = \frac{DCG@k}{IDCG@k} = \frac{\sum_{i=1}^{k}\frac{rel_i}{\log_2(i + 1)}}{\sum_{i=1}^{k}\frac{rel_i^*}{\log_2(i + 1)}}
\end{equation}
where $rel_i$ is the relevance at position $i$ and $rel_i^*$ represents the ideal ranking. For theoretical foundations, see~\cite{ndcg_theory_wang2013}.
\\
\\
\noindent \textbf{Rank-Biased Precision (RBP)} models user browsing behavior with a persistence parameter $p$:
\begin{equation}
RBP = (1-p) \sum_{i=1}^{k} p^{i-1} \cdot rel_i
\end{equation}
where $p$ (typically 0.8) represents the probability that a user examines the next item.
\\
\\
\noindent \textbf{Mean Reciprocal Rank (RecipRank)} emphasizes the position of the first relevant item:
\begin{equation}
RecipRank = \frac{1}{|U|} \sum_{u=1}^{|U|} \frac{1}{rank_u}
\end{equation}
where $rank_u$ is the position of the first relevant item for user $u$.

\chapter{Related Work}

This chapter reviews existing research on ensemble methods in recommender systems and their environmental impact. We organize the literature into four main areas: ensemble methods in recommender systems, the environmental impact of machine learning and recommender systems, energy measurement methodologies, and the broader Green AI movement. Our analysis reveals a critical gap: while ensemble methods are known to improve accuracy and machine learning's environmental cost is increasingly studied, to our knowledge no prior work has measured the energy consumption of ensemble techniques in recommender systems. This gap motivates our research.

\section{Ensemble Methods in Recommender Systems}

\paragraph{Ensemble Strategies and Empirical Results}

Recent work demonstrates substantial accuracy improvements from ensemble methods. \citet{ensemble_greedy_mehta2024} proposed a greedy ensemble approach that merges top-k lists from ten base models across five datasets. Their method achieves more than 20\% average NDCG improvement over the single best model. Moreover, their greedy selection strategy, which iteratively adds models only if they improve validation performance, outperforms ensembles that include all available models. This suggests that selective ensemble construction is crucial for maximizing accuracy gains.
\\
\\
\noindent \citet{ensemble_graph_embedding_forouzandeh2020} explored ensemble techniques with graph-based models on MovieLens, reporting quality gains compared to individual models. Across the reviewed ensemble literature, common strategies include: (1) averaging ensembles that compute the mean of predictions, (2) weighted ensembles that assign learned weights to base models, (3) stacking ensembles that use meta-learners to combine predictions, and (4) selective ensembles that choose optimal model subsets.
\\
\\
\noindent Ensemble success relies on two key factors: diversity and competence. Models should make different errors so that combining their predictions cancels out mistakes. However, each individual model must still perform reasonably well. This relates to the bias-variance trade-off: ensembles reduce variance by averaging predictions while keeping bias low.

\paragraph{The Missing Environmental Perspective}

A systematic analysis of the reviewed ensemble research reveals a pattern: evaluations focus mostly on accuracy metrics (NDCG, precision, recall). Energy consumption, computational cost, and carbon footprint are not commonly reported in ensemble studies in recommender systems. \citet{ensemble_accuracy_energy_omar2024} highlight that adding more models to ensembles does not always justify the energy cost in general machine learning systems, but no work has investigated this trade-off compared to single optimized models in recommender systems.

\section{Environmental Impact of Machine Learning and Recommender Systems}

The environmental cost of machine learning has received growing attention, with recommender systems as a particular focus due to their deployment at massive scale.

\paragraph{Energy Consumption in Recommender Systems}

\citet{recsys_carbon_footprint_vente2024} provided one of the first systematic measurements in terms of energy consumption in recommender systems research. They reproduced typical RecSys experimental pipelines and measured energy using hardware meters. Their results reveal dramatic differences between algorithm families: deep learning models consumed 1.79 kWh on average, while traditional UserKNN consumed only 0.03 kWh—approximately 60× less energy—despite achieving similar NDCG@10 performance. Traditional models achieve 95-98\% of deep learning accuracy while consuming only 10-20\% of the energy, showing diminishing returns. The final percentage points of accuracy improvements require disproportionate energy.
\\
\\
\noindent Scaling these measurements to full experimental papers reveals substantial environmental impact. A typical deep learning-based RecSys paper generates 3,297 kg CO$_2$ equivalents, exceeding emissions from a New York to Melbourne flight. Deep learning papers produce 42× more emissions than traditional method papers ~\citep{recsys_carbon_footprint_vente2024}.

\paragraph{Energy-Efficient Recommendations Research}

Beyond measuring consumption, recent work explores reducing energy consumption. \citet{dataset_size_energy_arabzadeh2024} investigated optimizing dataset size for energy efficiency. Smaller training datasets reduce computational requirements, but dataset size affects model accuracy. Finding the optimal balance between energy efficiency and accuracy is crucial.
\\
\\
\noindent \citet{gnn_sustainable_purificato2024} proposed eco-aware graph neural networks that balance recommendation quality with energy consumption, demonstrating that pipeline and structure choices significantly impact environmental impact. \citet{ensemble_model_selection_nijkamp2025} showed that strategic model selection for ensembles in production environments can reduce energy consumption while maintaining accuracy, though not specifically for recommender systems.
\\
\\
\noindent \citet{carbon_aware_recsys_kalisvaart2025} addressed sustainability within recommendation content itself. They created the RecipeEmission dataset where each recipe has an estimated CO$_2$ footprint. Benchmarking nine algorithms revealed that conventional models ignore item greenness. They proposed reranking based on combined accuracy-greenness scores, improving environmental outcomes with minimal accuracy loss.

\paragraph{Reproducibility and Experimental Rigor}

\citet{random_seeds_wegmeth2024} investigated random seed effects on recommender system evaluations. Testing 20 different random seeds for data splitting, they found accuracy varied by up to 6.3\% depending on seed choice. This variability is rarely reported in publications. Cross-validation reduces variability compared to single holdout splits. The work highlights the importance of experimental methodology for reproducible research.

\section{Energy Measurement Methodologies}

Accurate energy measurement is essential for evaluating the environmental impact of recommender systems. \citet{emers_tool_wegmeth2024} developed EMERS, an energy meter specifically designed for recommender systems research. The tool integrates hardware meters with experimental pipelines to enable reproducible energy measurements~\citep{emers_github}.

\paragraph{Measurement Approaches and Trade-offs}

\citet{power_meters_comparison_jay2023} compared software-based power meters, evaluating RAPL\footnote{Running Average Power Limit}, NVML\footnote{NVIDIA Management Library}, and other tools. Results showed reasonable accuracy for many tools, though significant deviations occurred in some scenarios. The choice of measurement tool affects results. Energy measurement itself introduces overhead: frequent polling consumes CPU cycles that may affect the measured workload.

\section{Green AI and Sustainable Computing}

The Green AI movement addresses the environmental cost of artificial intelligence research and deployment. \citet{green_ai_schwartz2019} raises awareness about this issue and proposed that papers should report efficiency metrics—accuracy-per-watt or accuracy-per-CO$_2$—alongside traditional accuracy metrics. This enables researchers to make informed trade-offs: a model with 1\% better accuracy but 10× higher energy consumption may not be desirable.

\paragraph{Carbon Footprint Accounting}

\citet{computing_carbon_gupta2021} investigated the comprehensive environmental footprint of computing, identifying these two main components: operational carbon emissions from electricity consumption during use, and embodied carbon emissions from manufacturing. Hardware production requires energy and rare earth elements. Mining these elements also has environmental costs including water pollution and habitat destruction.
\\
\\
\noindent Carbon footprint itself varies significantly by region of deployment. France has low carbon intensity due to nuclear power, while Poland has high intensity due to coal. Geographic location and time of day, which affects renewable energy availability, significantly impacts environmental cost. \citet{ml_energy_reporting_henderson2020} called for systematic lifecycle assessment in machine learning research, covering manufacturing, operation, and disposal.

\paragraph{Best Practices and Research Context}

\citet{recsys_sustainability_overview_felfernig2023} identified research challenges in sustainable recommender systems. \citet{recsys_sustainability_tradeoff_spillo2023} emphasized the need for multi-objective optimization balancing algorithm performance and carbon footprint, which is central to our investigation of ensemble methods.

\section{Research Gap and Thesis Positioning}

 Our literature review reveals that two established research areas have not yet intersected sufficiently. Ensemble methods consistently improve recommender system accuracy through diverse strategies: averaging, weighting, stacking, selective ensembles. Simultaneously, environmental impact research has quantified substantial energy differences between algorithm families. However, to our knowledge no prior work has measured the energy consumption of ensemble techniques compared to single optimised models in recommender systems.
\\
\\
\noindent Important questions remain unanswered: What is the energy cost of achieving ensemble accuracy gains? How does ensemble overhead scale with dataset characteristics? Which ensemble strategies are more energy-efficient? These questions are practically important: recommender systems operate at massive scale where even small efficiency improvements translate to significant environmental impact.
\\
\\
\noindent This thesis addresses this gap through systematic evaluation. We measure energy consumption of ensemble techniques using hardware meters (EMERS) across two experimental pipelines (Surprise for explicit rating prediction, LensKit for implicit feedback ranking), four datasets (MovieLens 100K/1M, ModCloth, Anime), and four ensemble strategies (Average, Weighted, Stacking, Top Performers). We compare ensembles against optimized single models to determine when ensembles justify their environmental cost.

\chapter{Method}

This study investigates the research question \textit{"How do ensemble techniques in recommender systems influence the environmental impact compared to a single optimized model?"}. The research employs a systematic experimental approach to quantify the trade-offs between prediction accuracy and energy consumption across different recommendation algorithms.
\\
\\
\noindent The methodology consists of several key steps: dataset selection, data preprocessing, model implementation, energy consumption measurement, and quality assurance tests. Models are categorized into three groups: simple baseline models, single optimized models, and ensemble models. All models are evaluated on four benchmark datasets across different domains and scales. Prediction accuracy is measured by RMSE for explicit rating prediction and NDCG@10 for ranking-based recommendation. Energy consumption is measured using a smart plug that monitors complete system power draw during training and prediction phases, with environmental cost reported in watt-hours (Wh) and carbon emissions (gCO\textsubscript{2}e).

\section{Data Sets}
\label{sec:DataSets}

This study utilizes four datasets: \textbf{Anime}, \textbf{ModCloth}, \textbf{ML-100K}, and \textbf{ML-1M}. The MovieLens datasets (ML-100K and ML-1M) are well-established benchmarks in recommender systems research~\citep{movielens_dataset_harper2015}. These datasets are widely used in the research community. This enables comparison with multiple previous studies.
\\
\\
\noindent The datasets are obtained from a Google Drive folder~\citep{recbole_datasets} provided by RecBole~\citep{recbole_framework_zhao2021}. The folder contains 46 preprocessed datasets. These datasets are ready for direct use in the experiment pipeline. The preprocessing follows standardized procedures. This ensures consistent formatting across all datasets.
\\
\\
\noindent The selection includes datasets of different sizes and characteristics. ML-100K is a small dataset with 100,000 ratings. It is used for initial experimentation and configuration testing. ML-1M contains 1 million ratings. It provides a larger scale for evaluation. ModCloth is a fashion dataset with 82,790 interactions. Anime is a large dataset with 7.8 million interactions. These larger datasets enable more robust and meaningful evaluation.
\\
\\
\noindent The diverse dataset selection is important. MovieLens datasets represent movie ratings, ModCloth represents fashion preferences, and Anime represents entertainment preferences. This selection enables assessment of algorithm generalizability across different recommendation contexts. 

\begin{table}[htbp]
    \centering
    \caption{Overview of the four datasets used in this study, showing their characteristics and scale differences.}
    \label{tab:dataset_overview}
    \vspace{3mm}
    \begin{tabular}{lrrrrl}
        \hline
        \textbf{Dataset} & \textbf{Users} & \textbf{Items} & \textbf{Ratings} & \textbf{Sparsity} & \textbf{Rating Scale} \\
        \hline
        ML-100K & 943 & 1,682 & 100,000 & 93.70\% & 1-5 \\
        ML-1M & 6,040 & 3,706 & 1,000,209 & 95.53\% & 1-5 \\
        ModCloth\textsuperscript{a} & 47,958 & 1,378 & 82,790 & 99.87\% & 1-5 \\
        Anime\textsuperscript{b} & 73,515 & 11,200 & 7,813,737 & 99.05\% & 1-10 \\
        \hline
    \end{tabular}
    
    \vspace{0.3cm}
    \footnotesize
    \textsuperscript{a}Original scale is 0-5; preprocessed data contains only ratings 1-5.\\
    \textsuperscript{b}Original scale includes -1 for "watched but not rated"; these entries are excluded during preprocessing.
\end{table}

\section{Pre-processing and Data Splitting}
\label{sec:PreprocessingDataSplitting}

All recommendation algorithms and datasets are evaluated using a unified preprocessing and data splitting workflow. This ensures consistent prerequisites across all experiments. The experiment uses the following standardized preprocessing pipeline.

\paragraph{Data Preprocessing}

Each dataset is processed through a standardized preprocessing loader. The loader handles different attributes and formats specific to each dataset. 
\\
\\
\noindent The first step is \textbf{formatting normalization}. Incoming datasets are loaded into a consistent pandas DataFrame structure. The DataFrame uses standardized column names: \textit{user}, \textit{item}, and \textit{rating}. This standardization simplifies subsequent processing steps.
\\
\\
\noindent The second step is \textbf{data cleaning}. The dataset undergoes a multi-step cleaning process. First, duplicate entries are removed. Duplicate user-item pairs are identified. For datasets with timestamps (ML-100K, ML-1M, Anime), interactions are sorted chronologically and the most recent interaction is kept. For ModCloth, which lacks timestamp information, the last occurrence in the source file is retained. Second, missing values are removed. Any row containing NaN values is removed. 
\\
\\
\noindent The third step is \textbf{rating scale validation}. The data is filtered according to the specific rating scale of each dataset. ML-100K and ML-1M have 1-5 scales. ModCloth has ratings 1-5 in the preprocessed data (original scale 0-5). Anime has a 1-10 scale. Entries with ratings outside these ranges are removed. For Anime, ratings of -1 indicate "watched but not rated" and are excluded from the dataset.

\paragraph{Data Splitting Strategy}

After preprocessing, the datasets are ready for the splitting phase. We implement two distinct data splitting strategies. The choice of strategy depends on the evaluation task: explicit rating prediction or implicit ranking-based recommendation.

\subparagraph{Global Random Split (Surprise Pipeline)}
The global random split is used for explicit rating prediction experiments with the Surprise library. The entire dataset is split into training (80\%) and test sets (20\%) with random seed 0 for reproducibility. The split is cached in pickle files stored in \texttt{data/\{dataset\}-split/}. Each split directory contains \texttt{train.pkl} and \texttt{test.pkl}. If a split exists, it is loaded from cache. Otherwise, a new split is created and saved. This caching mechanism ensures all algorithms are evaluated on identical train-test splits.

\subparagraph{Per-User Random Split (LensKit Pipeline)}
The per-user random split is used for ranking-based recommendation experiments with the LensKit library. Unlike the global split, this method ensures that each user appears in both training and test sets.
\\
\\
\noindent The splitting process works as follows. Users with insufficient interactions are filtered out. A minimum threshold is applied based on dataset sparsity. For dense datasets (ML-100K and ML-1M), the minimum is 10 items per user. For moderately sparse datasets (Anime), the minimum is 5 items per user. For very sparse datasets (ModCloth), the minimum is 3 items per user. These thresholds were determined empirically. Initial experiments with a uniform threshold of 10 items worked well for ML-100K and ML-1M (less than 5\% user loss). However, ModCloth's extreme sparsity (99.87\%) resulted in 97\% user loss with the same threshold. The adaptive thresholds balance data quality. They result in sufficient items per user for meaningful evaluation with user coverage, while retaining enough users for statistical validity.
\\
\\
\noindent Then, for each remaining user, their interactions are shuffled randomly. This eliminates temporal bias and distribution shift that would occur with temporal splits. At the end, each user's interactions are split into training (80\%) and test portions (20\%). A minimum number of test items per user is enforced: 2 for ML-100K and ML-1M, 1 for Anime and ModCloth.

\paragraph{Cross-Validation}

In addition to the train-test split, 5-fold cross-validation is performed on the training set. Cross-validation provides robust performance estimates and validates model stability across different data partitions.
\\
\\
\noindent For the Surprise pipeline, cross-validation uses the built-in \texttt{cross\_validate} function from the Surprise library. The training data is partitioned into 5 folds, with RMSE calculated for each fold. The mean and standard deviation across folds are reported.
\\
\\
\noindent For the LensKit pipeline, cross-validation uses user-based folds. The \texttt{crossfold\_users} function creates 5 partitions with per-user splits. For each user, 20\% of their training items are held out for validation. Users with fewer than 5 interactions are excluded from cross-validation to ensure each fold has at least one test item with the 20\% test fraction. Ranking metrics (NDCG, RBP, RecipRank) are calculated for each fold, with mean and standard deviation reported.

\begin{figure}[H]
    \centering
    \includegraphics[width=\textwidth]{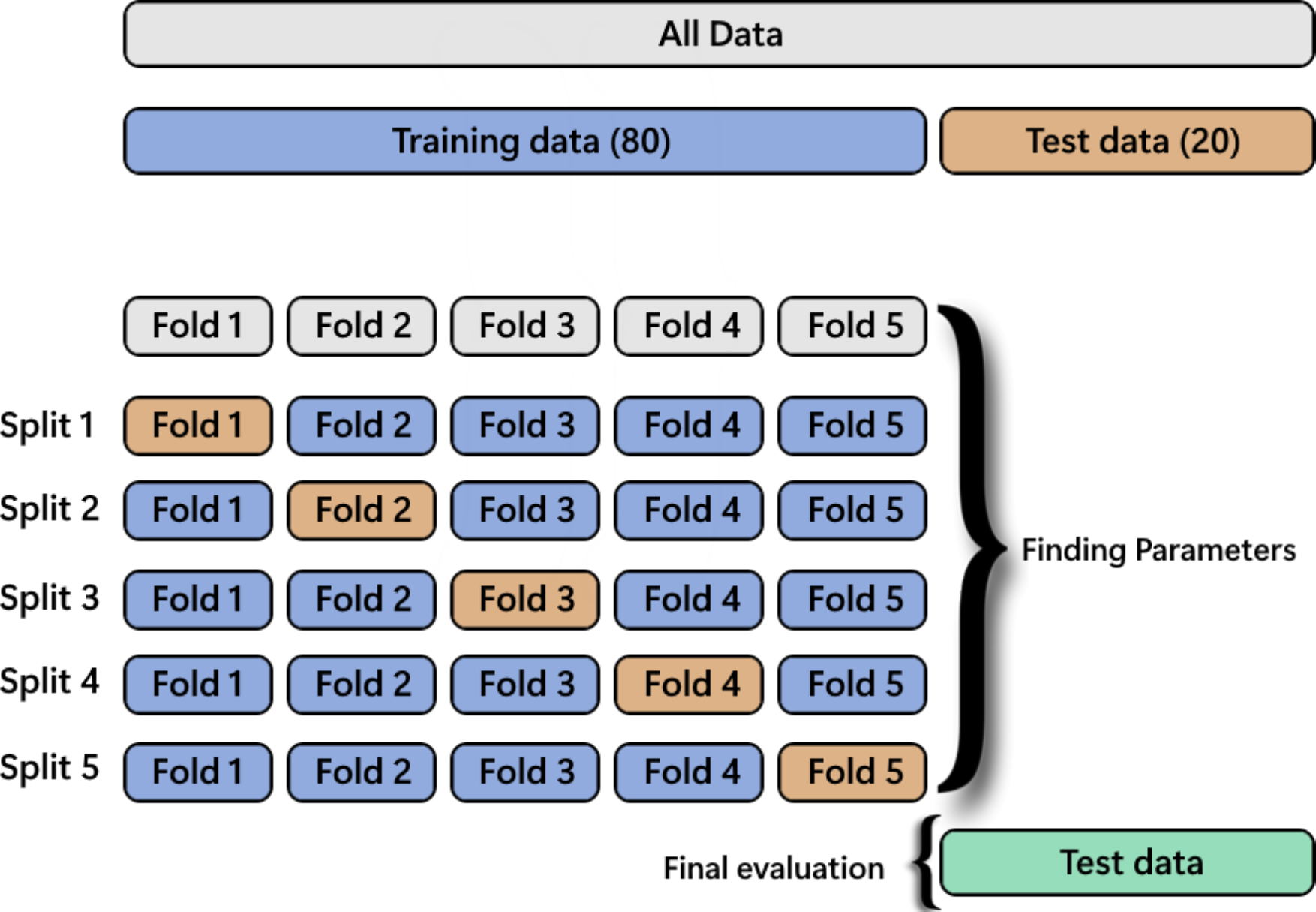}
    \caption{Illustration of the data splitting and cross-validation procedure. }
    \label{fig:split_cross_validation}
\end{figure}

\section{Model Selection}
\label{sec:ModelSelection}

The study compares three categories of recommendation models. This comparison reveals trade-offs between model complexity and predictive performance. The three categories are: simple baseline models, optimized single models, and ensemble models. Each category represents a different level of complexity and computational requirements. The Surprise pipeline evaluates 13 models (3 simple baselines, 6 optimized models, 4 ensemble methods). The LensKit pipeline evaluates 12 models (3 simple baselines, 5 optimized models, 4 ensemble methods). The goal is to keep both pipelines comparable.

\paragraph{Simple Baseline Models}

Simple baseline models provide a foundation for comparison. These models use minimal computational resources. They establish a performance baseline against which more complex models are evaluated. Three fundamental algorithms are included in this category.
\\
\\
\noindent The \textbf{Global Mean} predictor calculates the average rating across all interactions and predicts this global average for any user-item pair. The \textbf{Baseline Algorithm} from the Surprise library~\citep{surprise_joss_hug2020} models ratings as $r_{ui} = \mu + b_u + b_i$, accounting for user and item biases. The \textbf{Random Predictor} generates random predictions within the valid rating range, establishing a lower bound on performance. For the LensKit pipeline, three additional baselines are used: \textbf{Random Scorer} (random scores), \textbf{Popular Scorer} (ranks by global popularity), and \textbf{User Mean Scorer} (personalizes based on user activity level).

\paragraph{Optimized Single Models}

Optimized single models represent state-of-the-art approaches using advanced machine learning techniques. They require more computational resources but achieve significantly better prediction accuracy.
\\
\\
\noindent For the Surprise pipeline, six models are implemented. \textbf{Singular Value Decomposition (SVD)}~\citep{surprise_joss_hug2020} learns latent user and item factors through matrix factorization. \textbf{SVD++} extends SVD by incorporating implicit feedback. \textbf{Non-negative Matrix Factorization (NMF)} enforces non-negativity constraints for interpretability. \textbf{KNNBaseline} combines collaborative filtering with baseline predictions. \textbf{Slope One} computes average rating differences between item pairs. \textbf{Co-clustering} simultaneously clusters users and items to reduce sparsity.
\\
\\
\noindent For the LensKit pipeline, five models are implemented. \textbf{Alternating Least Squares (ALS)} learns user and item factors through alternating optimization for implicit feedback. \textbf{Bayesian Personalized Ranking (BPR)} uses pairwise learning to optimize for ranking. \textbf{Logistic Matrix Factorization} models interaction probability as a logistic function of user-item factors. \textbf{Item-Item K-Nearest Neighbors (Item-KNN)} and \textbf{User-User K-Nearest Neighbors (User-KNN)} build similarity matrices for items and users respectively.

\paragraph{Ensemble Models}

Ensemble models combine multiple algorithms to potentially improve predictive performance by capturing different aspects of user preferences (see Section~\ref{sec:EnsembleTechniques}). However, they require significantly more computational resources as they must train and execute multiple base models.
\\
\\
\noindent For the Surprise pipeline, four ensemble methods are implemented. \textbf{Average Ensemble} computes the arithmetic mean of predictions from SVD, SVD++, NMF, and KNNBaseline. \textbf{Weighted Ensemble} assigns optimized weights to the same base models based on validation performance. \textbf{Stacking Ensemble} uses a linear regression meta-learner to combine predictions from SVD, SVD++, NMF, and KNNBaseline. \textbf{Top Performers Ensemble} combines only SVD and SVD++, the best-performing models.
\\
\\
\noindent For the LensKit pipeline, four ensemble methods are implemented. \textbf{Average Ensemble} averages scores from ALS, BPR, Item-KNN, and User-KNN. \textbf{Weighted Ensemble} assigns learned weights to the same base rankers. \textbf{Rank Fusion} combines ranked lists from ALS, BPR, Item-KNN, and User-KNN using reciprocal rank fusion. \textbf{Top Performers Ensemble} combines only ALS and Item-KNN, the best-performing models.

\section{Experimental Pipelines}
\label{sec:ExperimentalPipelines}

We now have two separate pipelines with the same goal. The \textbf{Surprise pipeline} focuses on explicit rating prediction, treating recommendation as a regression task evaluated with RMSE (see Section~\ref{sec:EvaluationMetrics}). The \textbf{LensKit pipeline} focuses on implicit feedback and ranking-based recommendation, evaluated with ranking metrics (NDCG, RBP, RecipRank).
\\
\\
\noindent The Surprise pipeline uses the Surprise library~\citep{surprise_joss_hug2020} with global random splits. All models inherit from the \texttt{AlgoBase} class and implement standardized \texttt{fit} and \texttt{predict} methods. The LensKit pipeline converts explicit ratings to implicit feedback using threshold-based conversion (4.0 for 5-star scales, 7.0 for 10-star scales). It uses the LensKit framework with per-user splits to ensure each user appears in both training and test sets.
\\
\\
\noindent Both pipelines share common infrastructure: identical datasets, preprocessing procedures, data splitting (as described in Section~\ref{sec:PreprocessingDataSplitting}), cross-validation (5-fold), and energy measurement systems. Key differences include the splitting strategy (global vs. per-user), evaluation metrics (RMSE vs. ranking metrics), and data representation (explicit ratings vs. implicit binary feedback). Both pipelines measure energy consumption, training time, and prediction time to enable analysis of accuracy-efficiency trade-offs.

\begin{figure}[H]
    \centering
    \includegraphics[width=\textwidth]{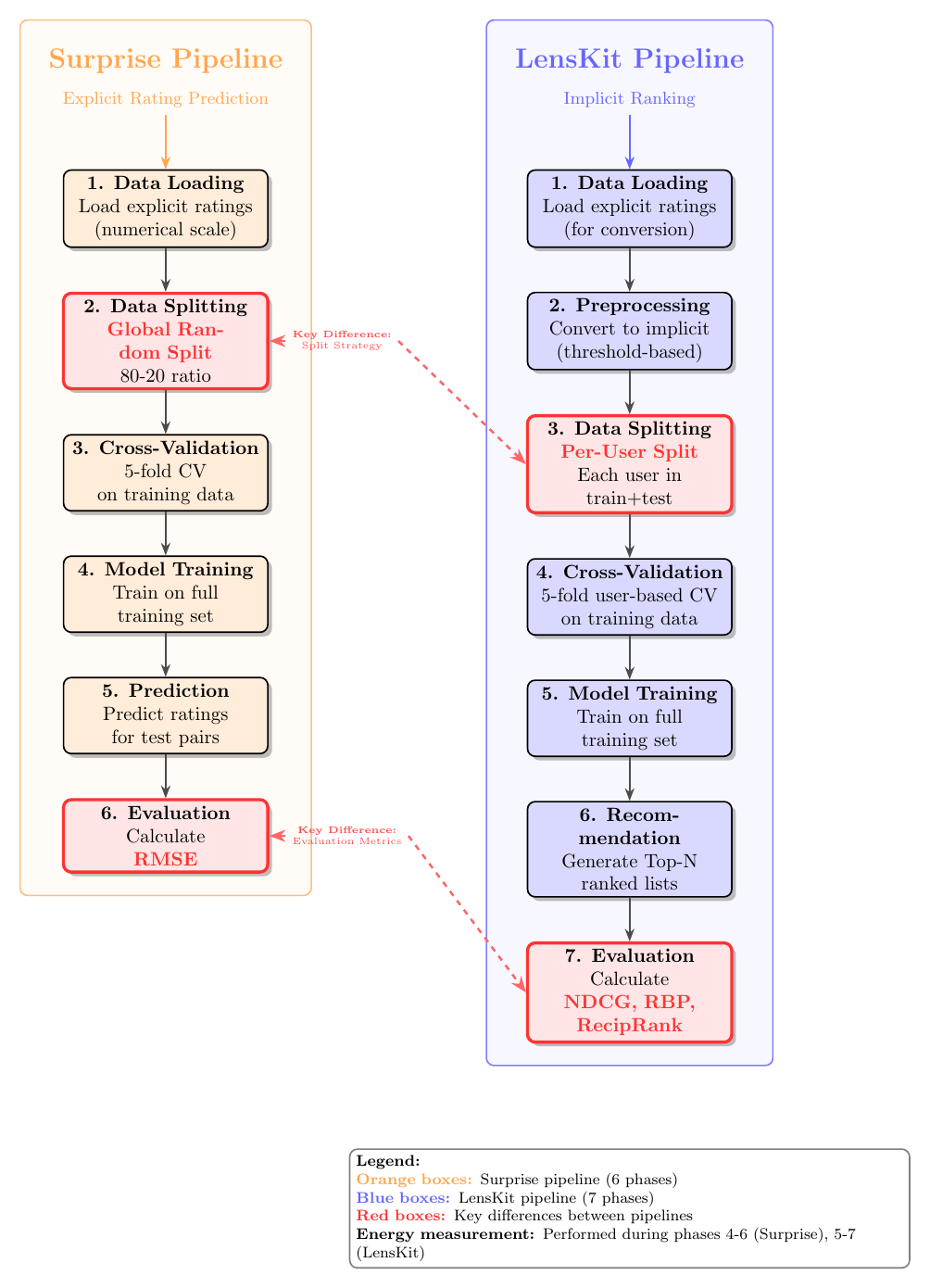}
    \caption{Comparison of the Surprise and LensKit experimental pipelines. The Surprise pipeline (left, orange) implements explicit rating prediction with 6 phases and the LensKit pipeline (right, green) implements implicit ranking-based recommendation with 7 phases. Red boxes highlight the key architectural differences.}
    \label{fig:pipeline_workflow}
\end{figure}

\section{Measuring Energy Consumption}
\label{sec:MeasuringEnergyConsumption}

Energy consumption measurement is a critical component of this study. The measurement uses EMERS (Energy Meter for Recommender Systems) \cite{emers_tool_wegmeth2024}. EMERS is integrated directly into the experimental pipeline. It enables real-time monitoring of energy usage during both training and prediction phases.

\paragraph{Hardware Setup and Configuration}

A Shelly Plug Plus S smart plug measures total system energy consumption during experiments. The device connects via WiFi, reporting power consumption (watts) and cumulative energy (watt-hours) at regular polling intervals. Device specifications: 0-2500W measurement range, ±1W accuracy, 0.1Wh resolution, 1Hz internal sampling, 230V AC operation.

\begin{figure}[H]
    \centering
    \includegraphics[width=\textwidth]{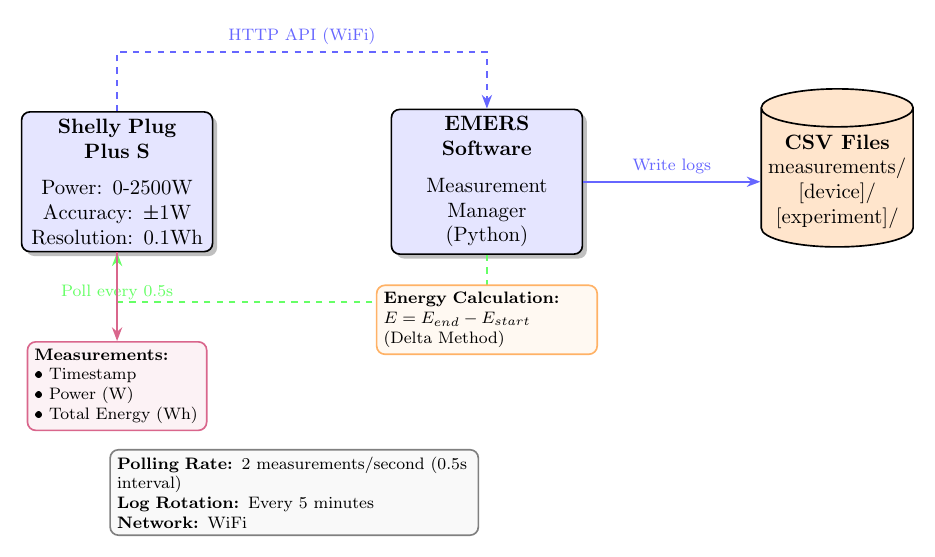}
    \caption{EMERS measurement setup architecture. The experimental computer's power consumption flows through the Shelly Plug Plus S smart plug, which measures total energy usage.}
    \label{fig:emers_setup}
\end{figure}

\paragraph{Integration with Experimental Pipeline}

Each experimental run uses the \texttt{Mea\-surementManager} class from EMERS via Python context managers. Upon entering the context, the manager loads device configuration from \texttt{emers/\-settings.json}, creates a unique experiment name in the format \texttt{EXPERIMENT\_\-[model]\_\-[dataset]\_\-[timestamp]}, and starts a background thread that polls the smart plug every 0.5 seconds. Each measurement records timestamp, current power draw (watts), and cumulative energy (watt-hours). Upon exit, the manager stops polling, closes log files, and releases resources.
\\
\\
\noindent Measurements are stored in experiment-specific directories with the structure \texttt{mea\-surements/\-[device]/\-[experiment\_\-name]/} and CSV log rotation every 5 minutes to manage file sizes and provide backup.

\paragraph{Energy Calculation and Data Processing}

Energy consumption is calculated using the delta method: $E_{experiment} = E_{end} - E_{start}$, where $E_{start}$ and $E_{end}$ are the cumulative energy values from the first and last measurements. This approach accounts for all energy consumption, is robust to temporary measurement failures, and handles log rotation automatically without accumulating numerical errors from integration.

\paragraph{Carbon Footprint Calculation}

Carbon emissions are calculated using grid-specific emission factors configured in \texttt{emers/\allowbreak monitor\_\allowbreak settings.json}. The conversion formula is: $C = (E_{Wh} / 1000) \times EF$, where $EF$ is the emission factor in gCO\textsubscript{2}e/kWh (e.g., 420 gCO\textsubscript{2}e/kWh for Germany, 275 gCO\textsubscript{2}e/kWh for European average). Results are reported in grams of CO\textsubscript{2} equivalent (gCO\textsubscript{2}e)~\cite{emers_tool_wegmeth2024}.

\section{Experimental Reliability and Quality Assurance}
\label{sec:ExperimentalReliability}

Scientific validity and reproducibility are essential for this study. Several methodological safeguards control for external factors that could influence measurements, ensuring that observed differences in energy consumption are attributable to the algorithms themselves.

\paragraph{Baseline Energy Consumption Measurement}

Baseline energy consumption is measured before running algorithm experiments to establish the idle power draw of the system. The system was left idle for 10 minutes with EMERS logging enabled but no experiments running, revealing an average baseline consumption of 71.2W (±5W).

\paragraph{Experimental Environment Control}

A standardized experimental environment is maintained throughout all experiments to minimize unrelated variables. All third-party applications are closed before running experiments. Unnecessary background processes are terminated. The hardware configuration remains identical throughout all experiments with no component changes, driver updates, or BIOS modifications.

\paragraph{Experimental Execution Protocol}

Each experiment follows a standardized protocol. The computer is restarted before each session to clear cached data and reset system state. Baseline power consumption is verified to be within acceptable range before proceeding. During execution, no user interaction occurs and the computer is dedicated solely to the experiment. Post-experiment validation checks energy consumption values for plausibility. Suspicious measurements are flagged and re-run if necessary. Data integrity checks verify dataset checksums, train-test split consistency, and model prediction ranges.

\paragraph{Reproducibility Measures}

All random number generation uses fixed seeds (random\_state=0 for both pipelines, np.random.seed(0) for NumPy operations). Train-test splits are cached to disk and reused for all subsequent experiments. Non-deterministic algorithms are configured for reproducibility using fixed random seeds. The complete experimental environment is documented in \texttt{requirements.txt}, including Python version 3.11, library versions, and dependencies. The entire code is available on GitHub~\citep{thesis_github_repo}.

\section{Hardware and Software Specifications}
\label{sec:HardwareSoftwareSpecs}

The experiments are conducted on a personal computer with specific hardware and software configurations. These specifications are documented to enable reproducibility.

\paragraph{Hardware Specifications}

The experimental computer has the following components:

\begin{itemize}
    \item \textbf{Operating System:} Windows 11 Pro, Version 24H2, Build 26100.4946
    \item \textbf{Processor (CPU):} AMD Ryzen 7 7800X3D (8 cores, 16 threads, 4.20 GHz base clock)
    \item \textbf{Memory (RAM):} 32GB DDR5 (6000 MHz, dual-channel)
    \item \textbf{Graphics Card (GPU):} AMD Radeon RX 7900 XTX, 24GB GDDR6
    \item \textbf{Power Consumption Meter:} Shelly Plug Plus S
\end{itemize}

\chapter{Results}

This chapter presents the empirical results of our experiments evaluating the environmental impact of ensemble techniques in recommender systems. We report findings from our two distinct recommendation pipelines: the Surprise framework for explicit rating prediction and the LensKit framework for implicit feedback ranking. For each framework, we present accuracy metrics, energy consumption, and carbon footprint measurements across four datasets: MovieLens-100K, MovieLens-1M, ModCloth, and Anime.

\section{Surprise Framework: Explicit Rating}

The Surprise framework evaluates explicit rating prediction, where users provide numerical ratings for items. We measured Root Mean Square Error (RMSE) as the accuracy metric, with lower values indicating better performance. Energy consumption was measured in watt-hours (Wh) and carbon emissions in grams of CO\textsubscript{2} equivalent.

\subsection{Comprehensive Model Comparison}

Table~\ref{tab:surprise_rmse_comparison} provides a comprehensive comparison of all Surprise models across datasets. The table shows RMSE scores for each dataset, the average RMSE across all datasets, and the relative performance compared to the optimized SVD model. Negative percentages indicate better performance than SVD, while positive percentages indicate worse performance. The ensemble method Top-Performers Ensemble achieves the best overall performance with 2\% improvement over SVD.

\begin{table}[H]
\centering
\caption{Comparison of RMSE scores of Individual model and Ensemble Method across various datasets}
\label{tab:surprise_rmse_comparison}
\vspace{3mm}
\resizebox{\textwidth}{!}{%
\begin{tabular}{l|cccc|c|c}
\hline
\textbf{Model} & \textbf{100K} & \textbf{1M} & \textbf{ModC.} & \textbf{Anime} & \textbf{Avg.} & \textbf{\% vs SVD} \\
\hline
\multicolumn{7}{l}{\textit{Simple Models}} \\
\hline
Baseline & 0.945 & 0.908 & \cellcolor{blue!25}0.942 & 1.206 & 1.000 & 3\% \\
Global Mean & 1.126 & 1.114 & 0.992 & 1.573 & 1.202 & 24\% \\
Random & 1.522 & 1.505 & 1.332 & 2.149 & 1.627 & 67\% \\
\hline
\multicolumn{7}{l}{\textit{Optimized Models}} \\
\hline
SVD & 0.935 & 0.872 & 0.946 & \cellcolor{blue!25}1.133 & 0.972 & 0\% \\
SVD++ & \cellcolor{blue!25}0.919 & \cellcolor{blue!25}0.862 & 0.944 & 1.192 & 0.979 & 1\% \\
NMF & 0.964 & 0.916 & 1.073 & 2.227 & 1.295 & 33\% \\
KNN-Baseline & 0.932 & 0.894 & 0.972 & ---$^*$ & --- & --- \\
Slope One & 0.947 & 0.906 & 1.052 & 1.197 & 1.025 & 6\% \\
Co Clustering & 0.967 & 0.915 & 1.017 & 1.209 & 1.027 & 6\% \\
\hline
\multicolumn{7}{l}{\textit{Ensemble Models}} \\
\hline
Average Ensemble & 0.917 & 0.863 & \cellcolor{green!25}0.936 & 1.168$^*$ & 0.971 & 0\% \\
Weighted Ensemble & 0.917 & 0.863 & 0.936 & 1.168$^*$ & 0.971 & 0\% \\
Stacking Ensemble & 0.918 & 0.860 & 0.941 & 1.224 & 0.986 & 1\% \\
Top-Performers Ens. & \cellcolor{green!25}0.915 & \cellcolor{green!25}0.854 & 0.938 & \cellcolor{green!25}1.120 & \cellcolor{green!25}0.957 & \cellcolor{green!25}-2\% \\
\hline
\end{tabular}%
}
\vspace{2mm}
\begin{flushleft}
\footnotesize
\textcolor{blue!70}{Blue highlighting} indicates best performing single model (Simple or Optimized) per dataset. \\
\textcolor{green!50!black}{Green highlighting} indicates best overall performance including ensemble models. \\
$^*$ Average and Weighted Ensemble on Anime dataset were executed without KNN-Baseline due to memory constraints (34.5 GB required for item-similarity matrix).
\end{flushleft}
\end{table}

\noindent Table~\ref{tab:surprise_energy_comparison} provides a comprehensive comparison of energy consumption across all Surprise models and datasets. The table shows energy consumption in milliwatt-hours (mWh) for each model-dataset combination, the average energy consumption across all datasets, and the percentage difference compared to SVD (the baseline optimized model). The simple models consume negligible energy on average, while ensemble methods show energy overheads ranging from +2,140\% to +2,549\% compared to SVD.

\begin{table}[H]
\centering
\caption{Comparison of Energy Consumption (mWh) of Individual Models and Ensemble Methods across various datasets}
\label{tab:surprise_energy_comparison}
\vspace{3mm}
\resizebox{\textwidth}{!}{%
\begin{tabular}{l|cccc|c|c}
\hline
\textbf{Model} & \textbf{100K} & \textbf{1M} & \textbf{ModC.} & \textbf{Anime} & \textbf{Avg.} & \textbf{\% vs SVD} \\
\hline
\multicolumn{7}{l}{\textit{Simple Models}} \\
\hline
Baseline & <0.0001 & 0.557 & 0.055 & 4.45 & 1.26 & -56\% \\
Global Mean & <0.0001 & 0.027 & 0.028 & <0.0001 & 0.014 & -100\% \\
Random & 0.035 & 0.430 & 0.027 & 3.29 & 0.945 & -67\% \\
\hline
\multicolumn{7}{l}{\textit{Optimized Models}} \\
\hline
SVD & 0.107 & 1.41 & 0.108 & 9.92 & 2.88 & +0\% \\
SVD++ & 1.82 & 38.9 & 0.117 & 181.7 & 55.6 & +1831\% \\
NMF & 0.061 & 1.95 & 0.347 & 13.5 & 3.97 & +38\% \\
KNN-Baseline & 0.302 & 17.9 & 3.12 & ---$^*$ & 7.11 & +147\% \\
Slope One & 0.178 & 7.26 & 0.054 & 33.5 & 10.2 & +254\% \\
Co Clustering & 0.089 & 2.26 & 0.365 & 15.6 & 4.58 & +59\% \\
\hline
\multicolumn{7}{l}{\textit{Ensemble Models}} \\
\hline
Average Ensemble & 3.13 & 70.0 & 4.27 & 227.2$^*$ & 76.1 & \cellcolor{red!25}+2542\% \\
Weighted Ensemble & 3.15 & 70.6 & 4.21 & 227.2$^*$ & 76.3 & \cellcolor{red!25}+2549\% \\
Stacking Ensemble & 2.77 & 50.0 & 1.31 & 227.2 & 70.3 & \cellcolor{red!25}+2341\% \\
Top-Performers Ens. & 2.79 & 46.2 & 0.817 & 208.4 & 64.5 & \cellcolor{red!25}+2140\% \\
\hline
\end{tabular}%
}
\vspace{2mm}
\begin{flushleft}
\footnotesize
Positive percentages indicate higher energy consumption than SVD, negative percentages indicate lower consumption. \\
\textcolor{red!70}{Red highlighting} indicates ensemble methods with extreme energy overhead. \\
$^*$Average and Weighted Ensemble on Anime dataset were executed without KNN-Baseline.
\end{flushleft}
\end{table}

\subsection{Accuracy-Energy Trade-offs}

\begin{figure}[H]
    \centering
    \includegraphics[width=\textwidth]{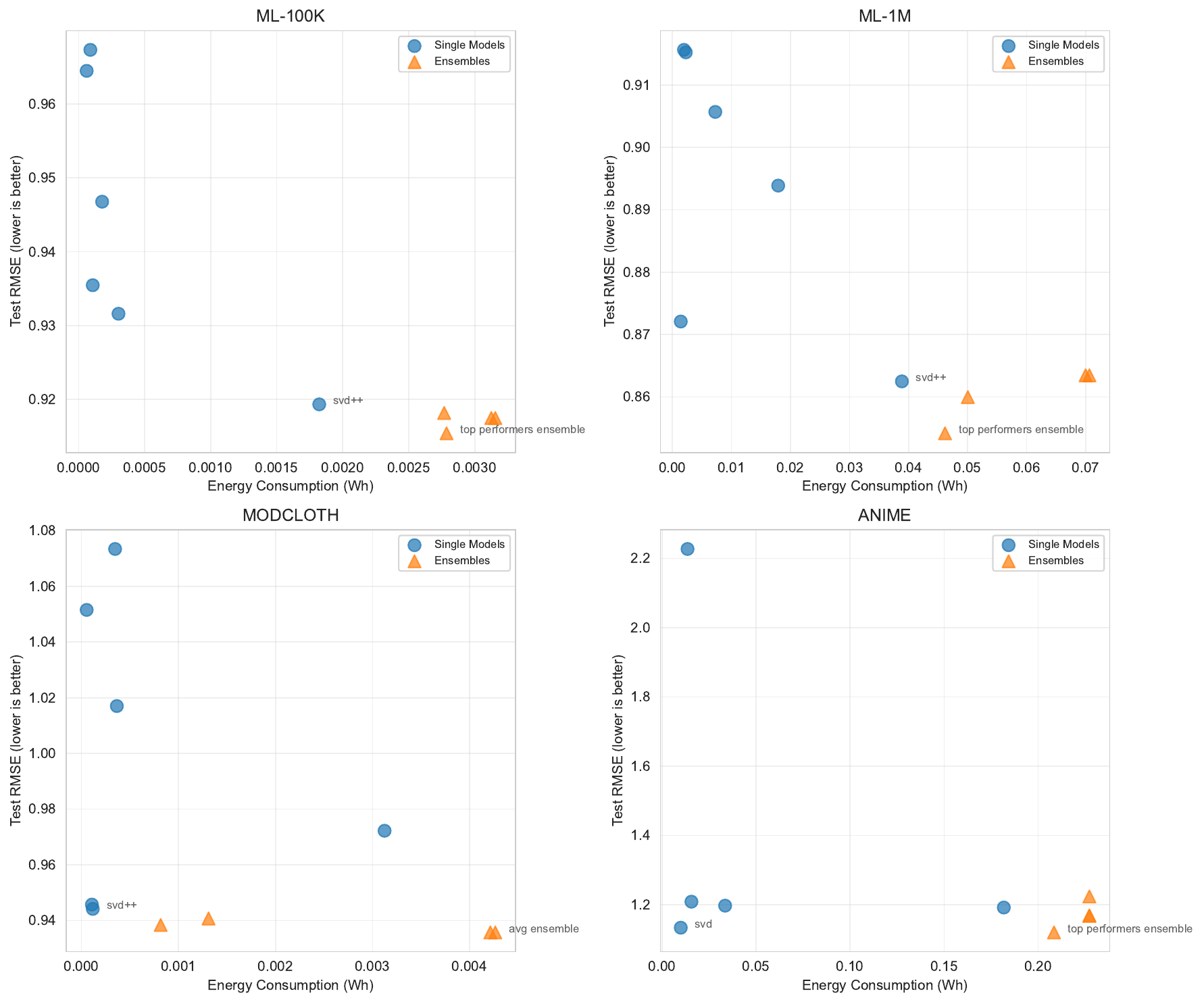}
    \caption{Accuracy vs. energy consumption for Surprise framework across all datasets. Single models (circles) and ensembles (triangles) show different trade-off patterns across dataset sizes. Anime dataset includes Average and Weighted ensembles without KNN-Baseline. Note the logarithmic scale for energy consumption.}
    \label{fig:surprise_scatter}
\end{figure}

Figure~\ref{fig:surprise_scatter} shows the relationship between energy consumption and accuracy across all datasets and model categories. The scatter plots reveal several patterns: on smaller datasets (ML-100K, ModCloth), most models cluster in similar accuracy ranges with relatively small energy differences. On larger datasets (ML-1M, Anime), both accuracy and energy consumption spread more widely, with ensemble methods consistently appearing in the higher energy consumption range while providing marginal accuracy improvements.

\begin{figure}[H]
    \centering
    \includegraphics[width=0.85\textwidth]{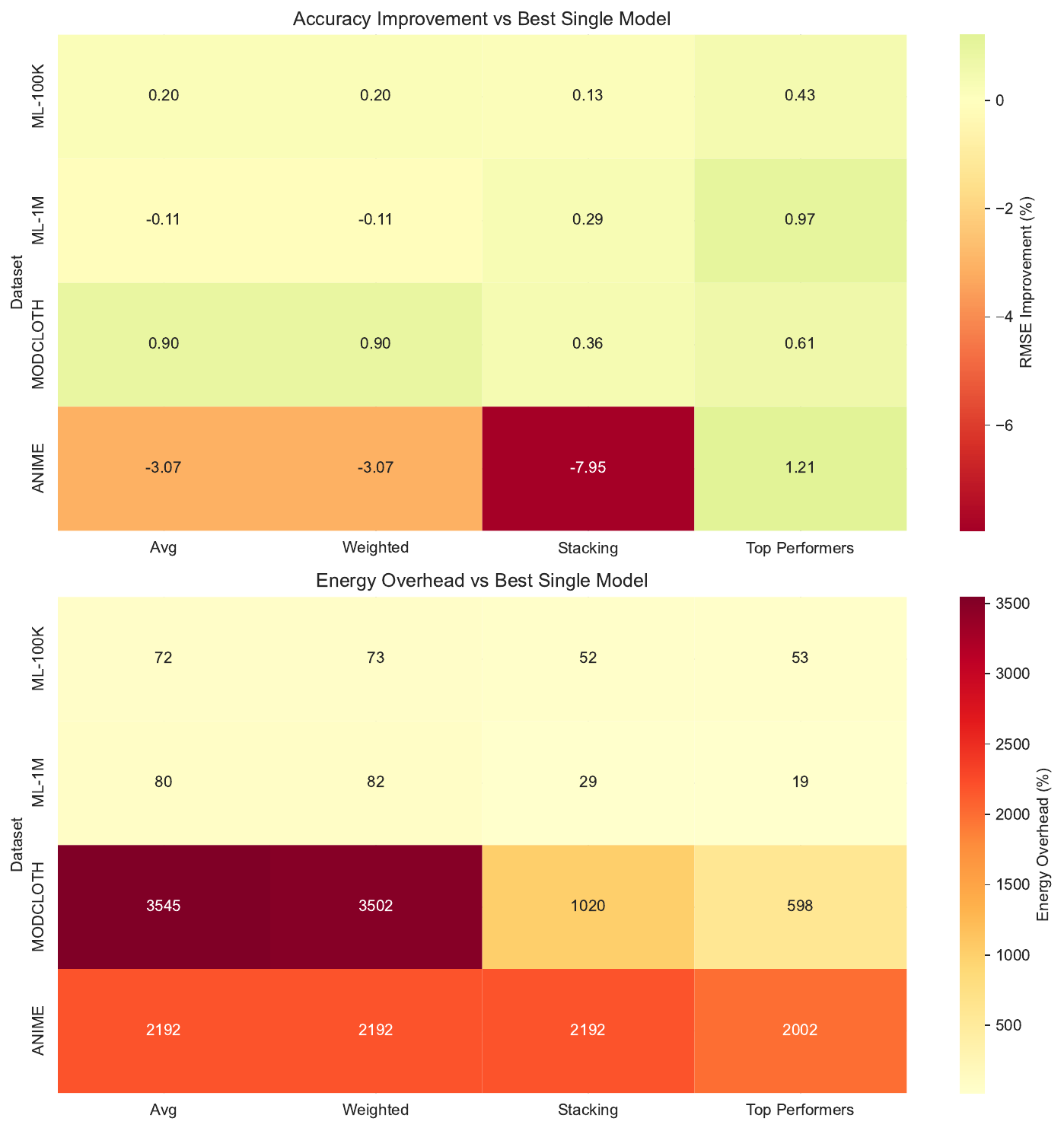}
    \caption{Ensemble efficiency heatmap for Surprise framework. Top: Accuracy improvement relative to best single model (positive values indicate better performance). Bottom: Energy overhead relative to best single model (higher values indicate more energy consumption). Anime dataset uses Average and Weighted ensembles without KNN-Baseline due to memory constraints.}
    \label{fig:surprise_heatmap}
\end{figure}

Figure~\ref{fig:surprise_heatmap} presents a heatmap visualization comparing ensemble strategies against the best single model for each dataset. The top panel shows accuracy improvement measured as percentage reduction in RMSE, while the bottom panel shows energy overhead showing percentage increase in energy consumption. Across all four datasets, accuracy improvements range from 0.1\% to 1.0\%, while energy overheads range from 19\% to 3,545\%, depending on the ensemble strategy and dataset.

\begin{figure}[H]
    \centering
    \includegraphics[width=\textwidth]{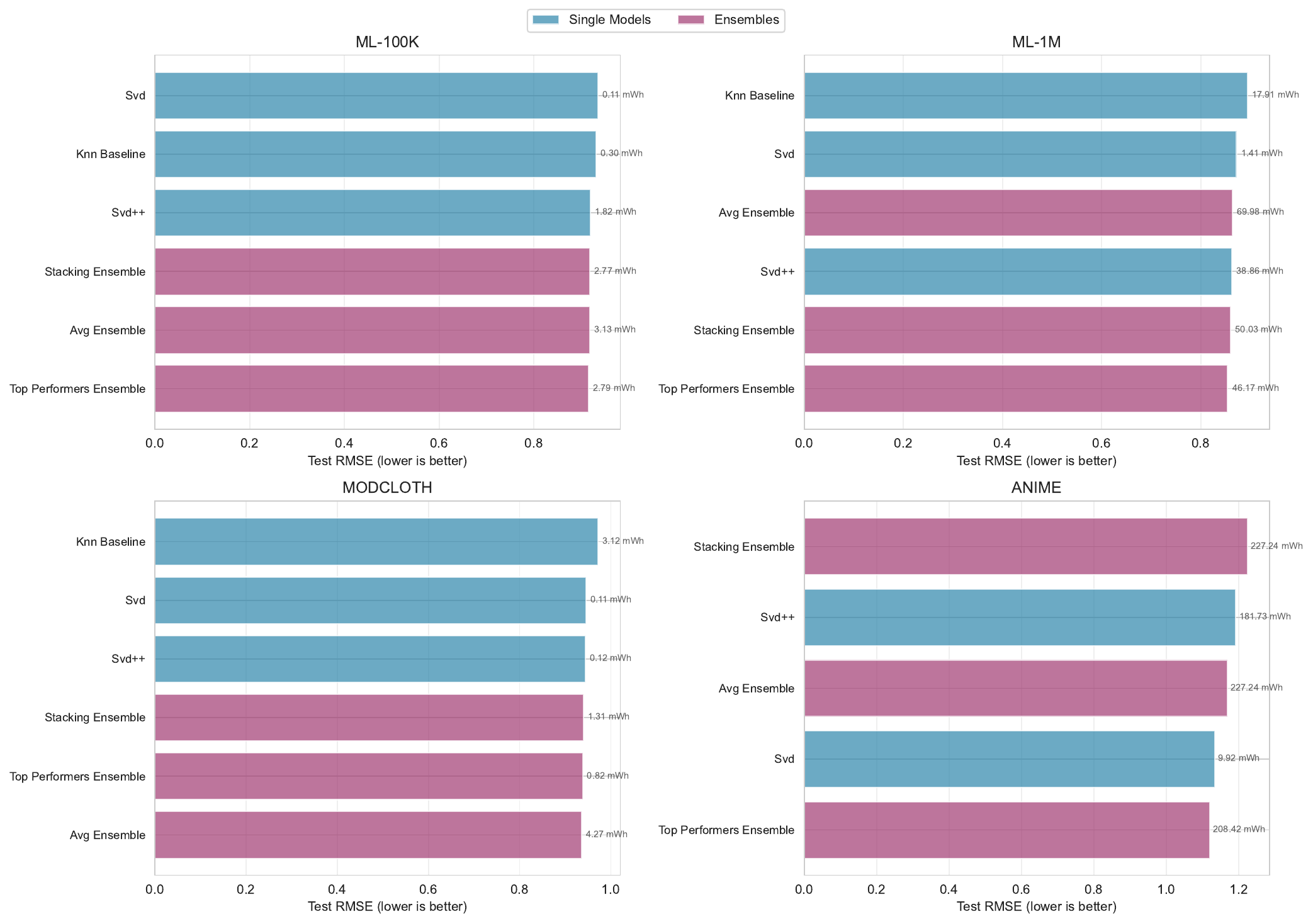}
    \caption{Model performance comparison for Surprise framework across all datasets. Bars show test RMSE (lower is better), with energy consumption (mWh) annotated. Blue bars represent single models, purple bars represent ensembles.}
    \label{fig:surprise_bars}
\end{figure}

\noindent Figure~\ref{fig:surprise_bars} presents a comparative bar chart visualization of model performance across datasets. The horizontal bars show test RMSE (shorter bars indicate better accuracy), with energy consumption displayed as text annotations. Single models are shown in blue, while ensemble methods appear in purple, making the accuracy-energy trade-offs immediately visible.

\subsection{MovieLens-100K Results}

On the MovieLens-100K dataset, SVD++ achieved the best test RMSE of 0.9193 among single models, consuming 0.0018 Wh of energy and producing 0.47 mg of CO\textsubscript{2}. The Top Performers ensemble, combining SVD++, SVD, and KNN Baseline, achieved a test RMSE of 0.9154, representing a 0.42\% improvement over SVD++. However, this improvement required 0.0028 Wh of energy, 52.9\% more than SVD++ alone. The Stacking ensemble achieved similar accuracy (RMSE: 0.9182) with comparable energy consumption (0.0028 Wh).
\\
\\
\noindent Among all ensemble strategies, the Average and Weighted ensembles showed nearly identical performance, both achieving test RMSE of 0.9175 with energy consumption of approximately 0.0031 Wh. The Baseline model, using only user and item biases, achieved a test RMSE of 0.9447 with negligible energy consumption.

\subsection{MovieLens-1M Results}

The MovieLens-1M dataset, showed larger differences between models. SVD++ achieved the best single-model test RMSE of 0.8625, consuming 0.0389 Wh and producing 10.03 mg of CO\textsubscript{2}. The Top Performers ensemble improved accuracy to 0.8542 RMSE, a 0.96\% improvement, while consuming 0.0462 Wh, which is 18.8\% more energy than SVD++.
\\
\\
\noindent The Stacking ensemble achieved 0.8600 RMSE with 0.0500 Wh energy consumption, while the Average and Weighted ensembles both achieved 0.8635 RMSE but consumed substantially more energy (0.0700 Wh and 0.0706 Wh respectively) due to longer prediction times. SVD, the second-best single model, achieved 0.8721 RMSE with only 0.0014 Wh, demonstrating significantly better energy efficiency than SVD++ while sacrificing 1.1\% accuracy.
\\
\\
\noindent The KNN Baseline model showed competitive accuracy (0.8939 RMSE) but required 0.0179 Wh due to expensive similarity computations during prediction. The Baseline model achieved 0.9078 RMSE with minimal energy consumption (0.00056 Wh).

\subsection{ModCloth Results}

On the ModCloth fashion e-commerce dataset, all optimized single models (SVD, SVD++, Baseline) achieved similar test RMSE values between 0.9420 and 0.9457, with the Baseline model performing competitively at 0.9420 RMSE. The Top Performers ensemble achieved the best accuracy among ensembles at 0.9384 RMSE, consuming 0.00082 Wh. 
\\
\\
\noindent The Average and Weighted ensembles achieved 0.9357 RMSE but required substantially more energy (0.0043 Wh and 0.0042 Wh respectively). Notably, the KNN Baseline model showed poor performance on this dataset, achieving only 0.9722 RMSE while consuming 0.0031 Wh.

\subsection{Anime Dataset Results}

The Anime dataset, containing 7.8 million ratings, presented unique challenges. SVD achieved the best single-model test RMSE of 1.1335, consuming 0.0099 Wh. SVD++, despite achieving training RMSE of 1.0314, performed worse on test data (1.1917 RMSE) while consuming 0.1817 Wh, indicating overfitting. 
\\
\\
\noindent The Top Performers ensemble achieved the best overall accuracy at 1.1197 RMSE, a 1.2\% improvement over SVD, but required 0.2084 Wh of energy—21 times more than SVD alone. The Stacking ensemble achieved 1.2236 RMSE with 0.2272 Wh consumption. Due to memory constraints, the KNN Baseline model could not be trained (requiring 34.5 GB for the item-similarity matrix), and consequently, the Average and Weighted ensembles also failed. The ensembles that did complete used only five base models instead of six.

\section{LensKit Framework: Implicit Feedback}

The LensKit framework evaluates implicit feedback ranking, where user preferences are inferred from behavioral signals. We measured NDCG@10 as the primary accuracy metric, with higher values indicating better performance. Rankings were generated from implicit feedback by converting ratings $\geq 4.0$ to positive interactions for all datasets except Anime, where the threshold was set to $\geq 7.0$ to account for its 1-10 rating scale.

\subsection{Comprehensive Model Comparison}

Table~\ref{tab:lenskit_ndcg_comparison} provides a comprehensive comparison of all LensKit models across datasets. The table shows NDCG@10 scores for each dataset, the average NDCG@10 across all datasets, and the relative performance compared to the best single model User-KNN. Positive percentages indicate better performance than User-KNN, while negative percentages indicate worse performance. All ensemble methods show comparable performance to User-KNN on average.

\begin{table}[H]
\centering
\caption{Comparison of NDCG@10 scores of Individual model and Ensemble Method across various datasets}
\label{tab:lenskit_ndcg_comparison}
\vspace{3mm}
\resizebox{\textwidth}{!}{%
\begin{tabular}{l|cccc|c|c}
\hline
\textbf{Model} & \textbf{100K} & \textbf{1M} & \textbf{ModC.} & \textbf{Anime} & \textbf{Avg.} & \textbf{\% vs User-KNN} \\
\hline
\multicolumn{7}{l}{\textit{Simple Models}} \\
\hline
Random & 0.006 & 0.004 & 0.009 & 0.001 & 0.005 & -98\% \\
Popular & 0.117 & 0.096 & 0.083 & 0.099 & 0.099 & -51\% \\
User Mean & 0.002 & 0.004 & 0.001 & 0.001 & 0.002 & -99\% \\
\hline
\multicolumn{7}{l}{\textit{Optimized Models}} \\
\hline
ALS & 0.157 & 0.155 & 0.127 & 0.202 & 0.160 & -21\% \\
BPR & 0.069 & 0.114 & 0.014 & 0.124 & 0.080 & -60\% \\
Logistic-MF & 0.074 & 0.116 & 0.008 & 0.137 & 0.084 & -59\% \\
Item-KNN & 0.194 & 0.154 & \cellcolor{blue!25}0.160 & 0.217 & 0.181 & -10\% \\
User-KNN & \cellcolor{blue!25}0.218 & \cellcolor{blue!25}0.196 & 0.159 & \cellcolor{blue!25}0.237 & \cellcolor{blue!25}0.202 & \cellcolor{blue!25}+0\% \\
\hline
\multicolumn{7}{l}{\textit{Ensemble Models}} \\
\hline
Average Ensemble & \cellcolor{green!25}0.230 & 0.203 & 0.162 & ---* & 0.198 & -2\% \\
Weighted Ensemble & 0.229 & 0.203 & \cellcolor{green!25}0.162 & ---* & 0.198 & -2\% \\
Rank Fusion & 0.226 & \cellcolor{green!25}0.207 & 0.161 & ---* & \cellcolor{green!25}0.198 & \cellcolor{green!25}-2\% \\
Top-Performers Ens. & 0.224 & 0.198 & 0.160 & ---* & 0.194 & -4\% \\
\hline
\end{tabular}%
}
\vspace{2mm}
\begin{flushleft}
\footnotesize
\textcolor{blue!70}{Blue highlighting} indicates best performing single model (Simple or Optimized) per dataset. \\
\textcolor{green!50!black}{Green highlighting} indicates best overall performance including ensemble models. \\
$^*$All ensemble methods failed on the Anime dataset due to memory constraints during prediction aggregation.
\end{flushleft}
\end{table}

\noindent Table~\ref{tab:lenskit_energy_comparison} provides a comprehensive comparison of energy consumption across all LensKit models and datasets. The table shows energy consumption in milliwatt-hours (mWh) for each model-dataset combination, the average energy consumption across the first three datasets (excluding Anime for consistency with ensemble results), and the percentage difference compared to User-KNN (the best single model). Ensemble methods show energy overheads ranging from +88\% to +263\% compared to User-KNN.

\begin{table}[H]
\centering
\caption{Comparison of Energy Consumption (mWh) of Individual Models and Ensemble Methods across various datasets}
\label{tab:lenskit_energy_comparison}
\vspace{3mm}
\resizebox{\textwidth}{!}{%
\begin{tabular}{l|cccc|c|c}
\hline
\textbf{Model} & \textbf{100K} & \textbf{1M} & \textbf{ModC.} & \cellcolor{gray!20}\textbf{Anime} & \textbf{Avg.$^{**}$} & \textbf{\% vs User-KNN} \\
\hline
\multicolumn{7}{l}{\textit{Simple Models}} \\
\hline
Random & 0.084 & 0.389 & 0.188 & \cellcolor{gray!20}18.9 & 0.220 & -76\% \\
Popular & 0.029 & 0.366 & 0.162 & \cellcolor{gray!20}21.6 & 0.186 & -80\% \\
User Mean & 0.058 & 0.307 & 0.163 & \cellcolor{gray!20}15.6 & 0.176 & -81\% \\
\hline
\multicolumn{7}{l}{\textit{Optimized Models}} \\
\hline
ALS & 0.062 & 0.661 & 0.196 & \cellcolor{gray!20}61.4 & 0.306 & -66\% \\
BPR & 1.17 & 11.0 & 0.331 & \cellcolor{gray!20}120.3 & 4.17 & +357\% \\
Logistic-MF & 1.01 & 11.3 & 0.304 & \cellcolor{gray!20}85.2 & 4.21 & +362\% \\
Item-KNN & 0.089 & 1.03 & 0.250 & \cellcolor{gray!20}47.6 & 0.456 & -50\% \\
User-KNN & 0.087 & 2.38 & 0.265 & \cellcolor{gray!20}177.7 & 0.911 & +0\% \\
\hline
\multicolumn{7}{l}{\textit{Ensemble Models}} \\
\hline
Average Ensemble & 0.177 & 4.48 & 0.468 & \cellcolor{gray!20}---$^*$ & 1.71 & +88\% \\
Weighted Ensemble & 0.179 & 4.89 & 0.441 & \cellcolor{gray!20}---$^*$ & 1.84 & +102\% \\
Rank Fusion & 0.340 & 8.89 & 0.703 & \cellcolor{gray!20}---$^*$ & 3.31 & +263\% \\
Top-Performers Ens. & 0.148 & 4.79 & 0.384 & \cellcolor{gray!20}---$^*$ & 1.77 & +94\% \\
\hline
\end{tabular}%
}
\vspace{2mm}
\begin{flushleft}
\footnotesize
Positive percentages indicate higher energy consumption than User-KNN, negative percentages indicate lower consumption. \\
$^*$ All ensemble methods failed on the Anime dataset due to memory constraints during prediction aggregation. \\
$^{**}$ Averages exclude the Anime dataset to ensure fair comparison with ensemble methods.
\end{flushleft}
\end{table}

\subsection{Accuracy-Energy Trade-offs}

\begin{figure}[H]
    \centering
    \includegraphics[width=\textwidth]{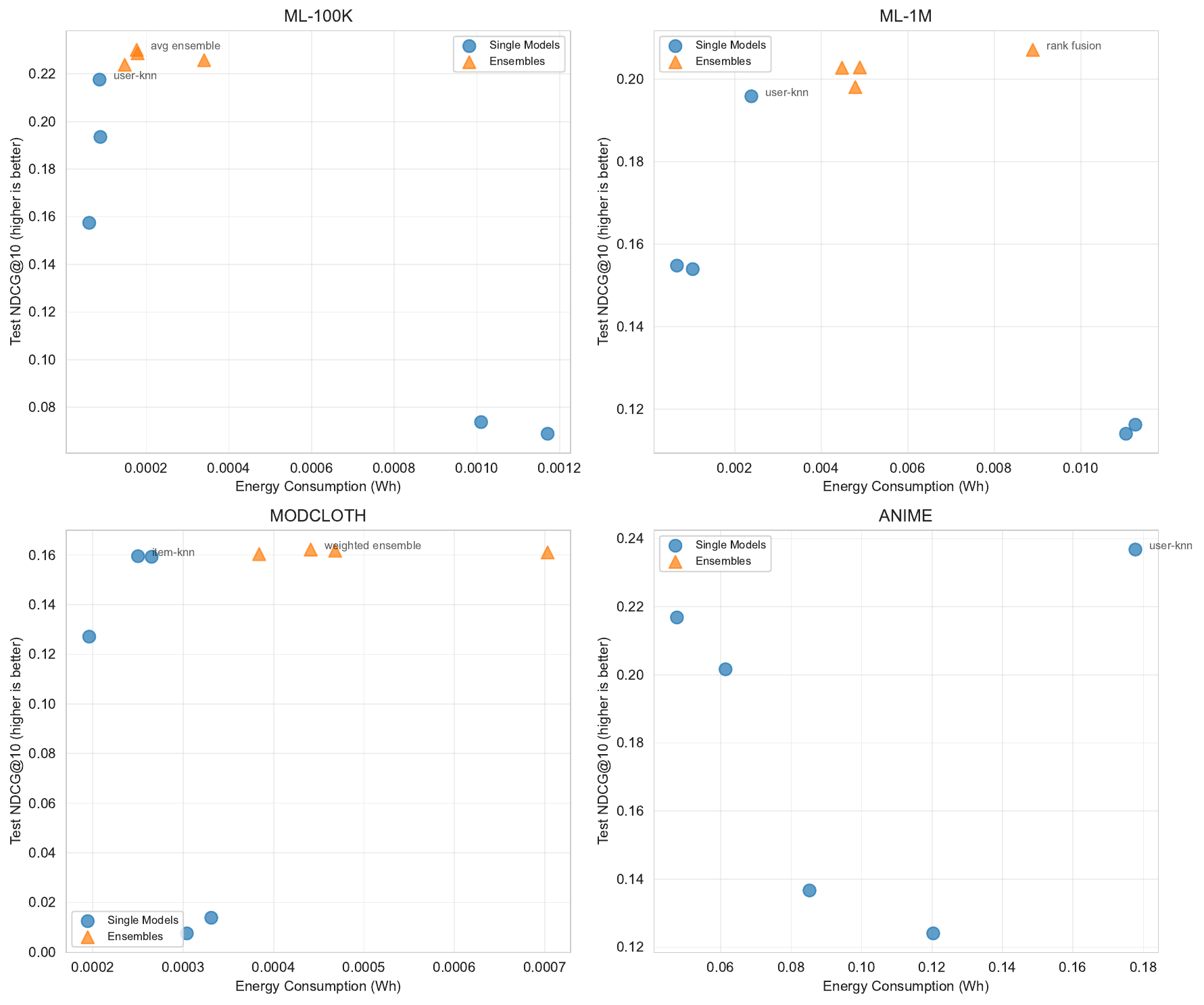}
    \caption{Accuracy vs. energy consumption for LensKit framework across all datasets. Single models (circles) and ensembles (triangles) demonstrate varying trade-off patterns. Note there are no ensemble results for the Anime dataset due to execution failures. Logarithmic scale used for energy consumption.}
    \label{fig:lenskit_scatter}
\end{figure}

\noindent Figure~\ref{fig:lenskit_scatter} illustrates the accuracy-energy relationship for LensKit models across all datasets. Similar to Surprise, smaller datasets show clustering of models in similar accuracy ranges with modest energy differences, while larger datasets exhibit greater spread in both dimensions. Ensemble methods consistently require higher energy consumption regions while providing small-to-moderate accuracy improvements.

\begin{figure}[H]
    \centering
    \includegraphics[width=0.85\textwidth]{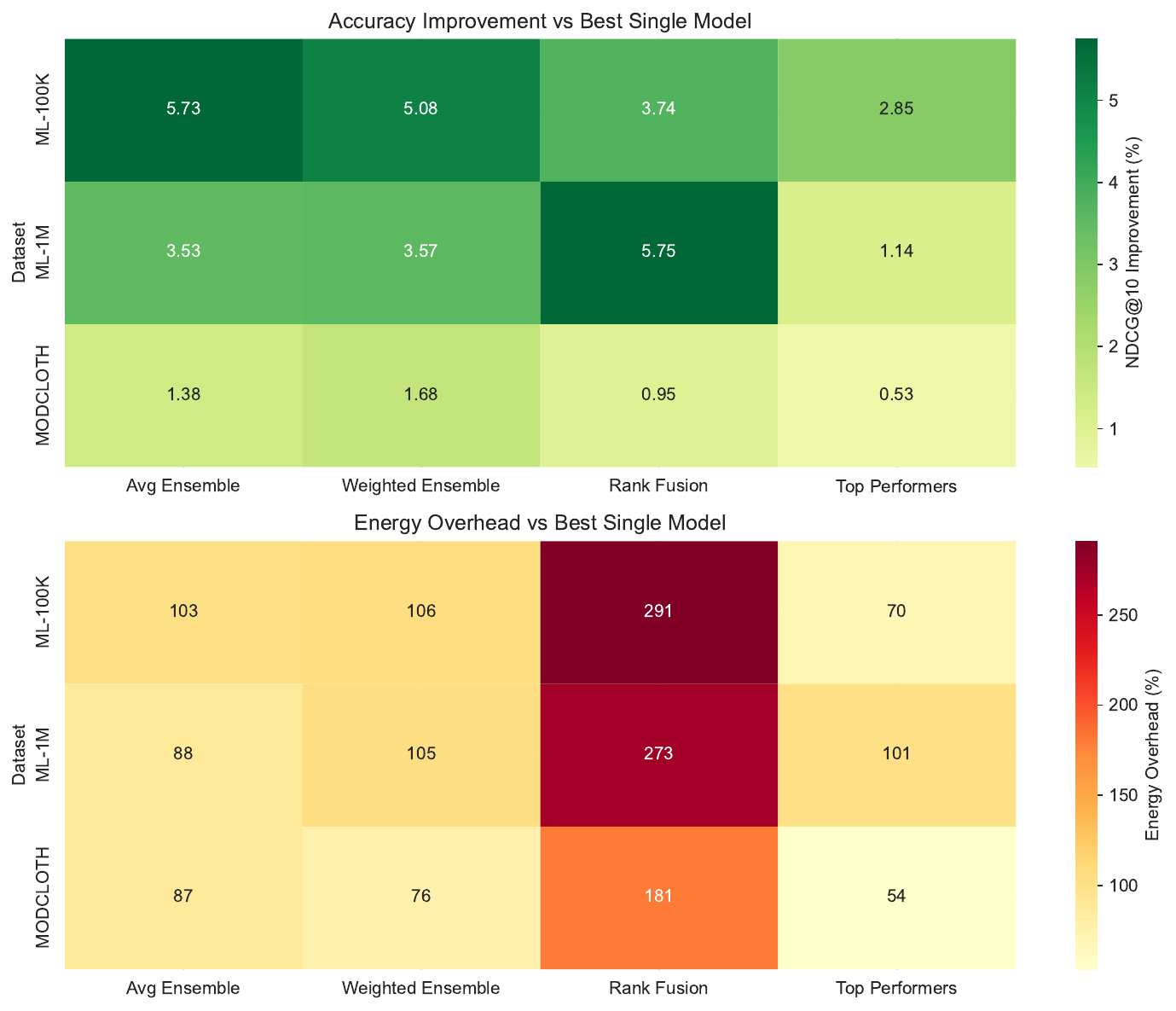}
    \caption{Ensemble efficiency heatmap for LensKit framework. Top: Accuracy improvement relative to best single model (positive values indicate better performance). Bottom: Energy overhead relative to best single model. Anime dataset excluded due to ensemble failures. Note there are no ensemble results for the Anime dataset due to execution failures.}
    \label{fig:lenskit_heatmap}
\end{figure}

\noindent Figure~\ref{fig:lenskit_heatmap} presents efficiency heatmaps comparing ensemble strategies against the best single model. Accuracy improvements (measured as percentage increase in NDCG@10) range from 1.1\% to 5.7\%, while energy overheads range from 76\% to over 700\%. The Rank Fusion strategy shows the highest energy overhead due to its computationally expensive score aggregation process.

\begin{figure}[H]
    \centering
    \includegraphics[width=\textwidth]{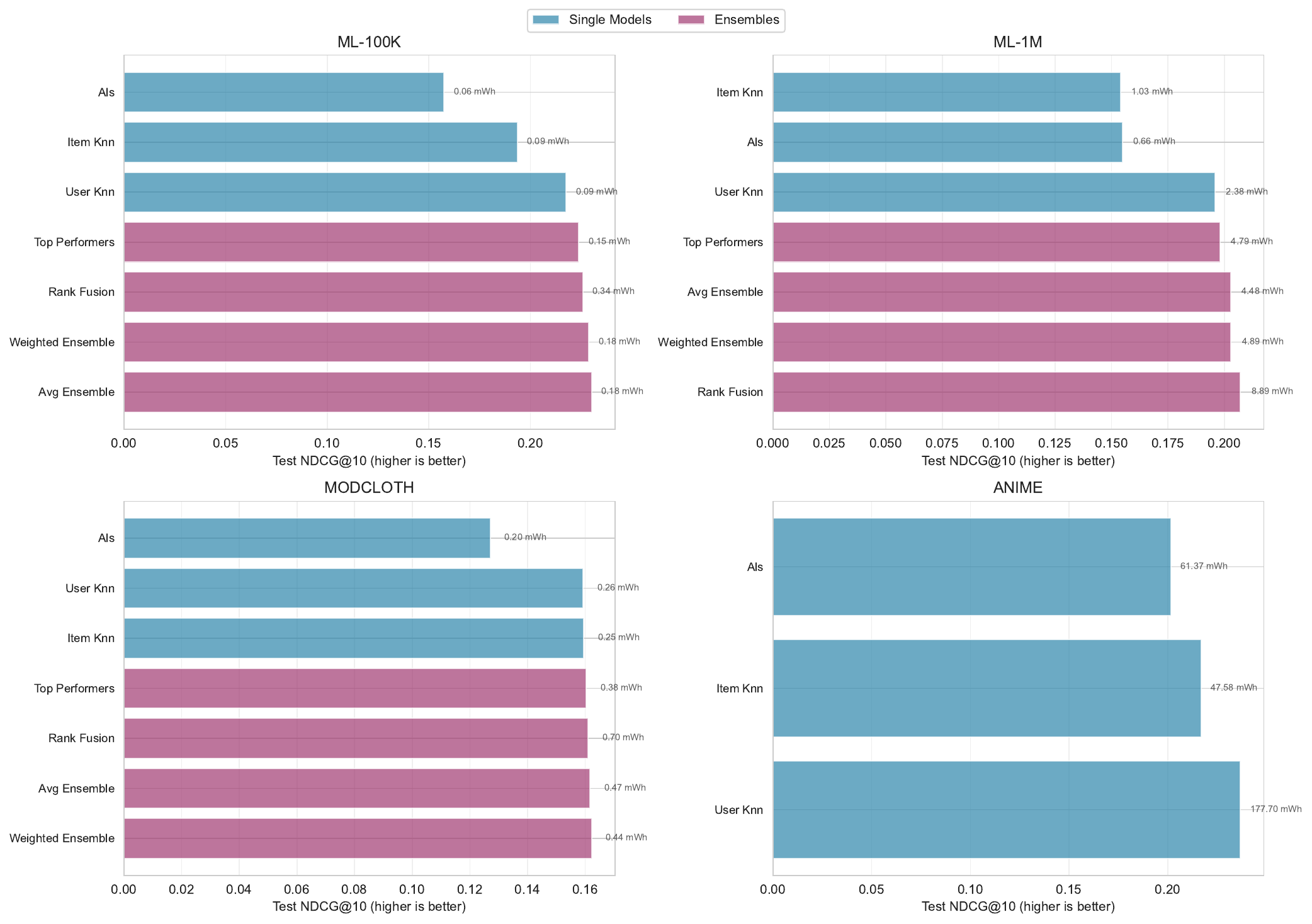}
    \caption{Model performance comparison for LensKit framework across all datasets. Bars show test NDCG@10 (higher is better), with energy consumption (mWh) annotated. Blue bars represent single models, purple bars represent ensembles. Anime dataset shows only single models due to ensemble memory errors.}
    \label{fig:lenskit_bars}
\end{figure}

\noindent Figure~\ref{fig:lenskit_bars} presents a comparative bar chart visualization of model performance across datasets. The horizontal bars show test NDCG@10 (longer bars indicate better accuracy), with energy consumption displayed as text annotations. Single models are shown in blue, while ensemble methods appear in purple. The visualization clearly illustrates that on smaller datasets, ensembles provide meaningful accuracy improvements at moderate energy costs, while on larger datasets like Anime, ensembles become infeasible due to memory constraints.

\subsection{MovieLens-100K Results}

On MovieLens-100K, User-KNN achieved the best single-model test NDCG@10 of 0.2176, consuming 0.000087 Wh. The Average ensemble improved accuracy to 0.2301 NDCG@10, a 5.7\% improvement, while consuming 0.00018 Wh, which is 103\% more energy than User-KNN. The Weighted ensemble achieved similar performance 0.2287 NDCG@10 with comparable energy consumption.
\\
\\
\noindent The Top Performers ensemble, combining ALS, BPR, and Logistic-MF, achieved 0.2238 NDCG@10 with 0.00015 Wh consumption, while the Rank Fusion ensemble achieved 0.2258 NDCG@10 with higher energy consumption 0.00034 Wh due to more complex score aggregation. Item-KNN achieved 0.1935 NDCG@10 with 0.000089 Wh, while Popular baseline achieved 0.1173 NDCG@10 with minimal energy (0.000029 Wh).

\subsection{MovieLens-1M Results}

On the larger MovieLens-1M dataset, User-KNN again performed best among single models with 0.1959 NDCG@10, consuming 0.0024 Wh. However, unlike the smaller dataset, the Rank Fusion ensemble achieved the highest overall accuracy at 0.2071 NDCG@10, a 5.7\% improvement over User-KNN. This improvement required 0.0089 Wh, representing a 270\% energy overhead.
\\
\\
\noindent The Weighted and Average ensembles achieved nearly identical performance (0.2029 and 0.2028 NDCG@10 respectively), with the Weighted ensemble consuming 0.0049 Wh and the Average ensemble consuming 0.0045 Wh. The Top Performers ensemble achieved 0.1981 NDCG@10 with 0.0048 Wh consumption. ALS, a matrix factorization approach optimized for implicit feedback, achieved 0.1548 NDCG@10 with only 0.00066 Wh, making it the most energy-efficient optimized model.

\subsection{ModCloth Results}

On the ModCloth dataset, the Weighted ensemble achieved the highest test NDCG\-@10 of 0.1622, consuming 0.00044 Wh. This represented a 1.3\% improvement over the best single model, Item-KNN (0.1595 NDCG@10, 0.00025 Wh). The Average ensemble achieved nearly identical performance (0.1617 NDCG@10) with slightly higher energy consumption (0.00047 Wh).
\\
\\
\noindent User-KNN achieved competitive accuracy 0.1593 NDCG@10 with 0.00027 Wh consumption, while ALS achieved 0.1271 NDCG@10 with 0.00020 Wh. The deep learning models BPR and Logistic-MF performed poorly on this dataset, achieving only 0.0138 and 0.0075 NDCG@10 respectively, despite consuming more energy (0.00033 Wh and 0.00030 Wh).

\subsection{Anime Dataset Results}

The Anime dataset presented severe scalability challenges for LensKit ensembles. User-KNN achieved the best single-model test NDCG@10 of 0.2368, but consumed 0.18 Wh due to the large user base. Item-KNN achieved 0.2168 NDCG@10 with 0.048 Wh, while ALS achieved 0.2016 NDCG@10 with 0.061 Wh.
\\
\\
\noindent All four ensemble strategies (Average, Weighted, Rank Fusion, and Top Performers) failed due to memory errors during prediction aggregation. The large number of users combined with ensemble score combination exceeded available system memory. The deep learning models completed execution but showed high energy consumption: BPR consumed 0.12 Wh for 0.1240 NDCG@10, while Logistic-MF consumed 0.085 Wh for 0.1366 NDCG@10.

\section{Cross-Framework Comparison}

\begin{figure}[H]
    \centering
    \includegraphics[width=\textwidth]{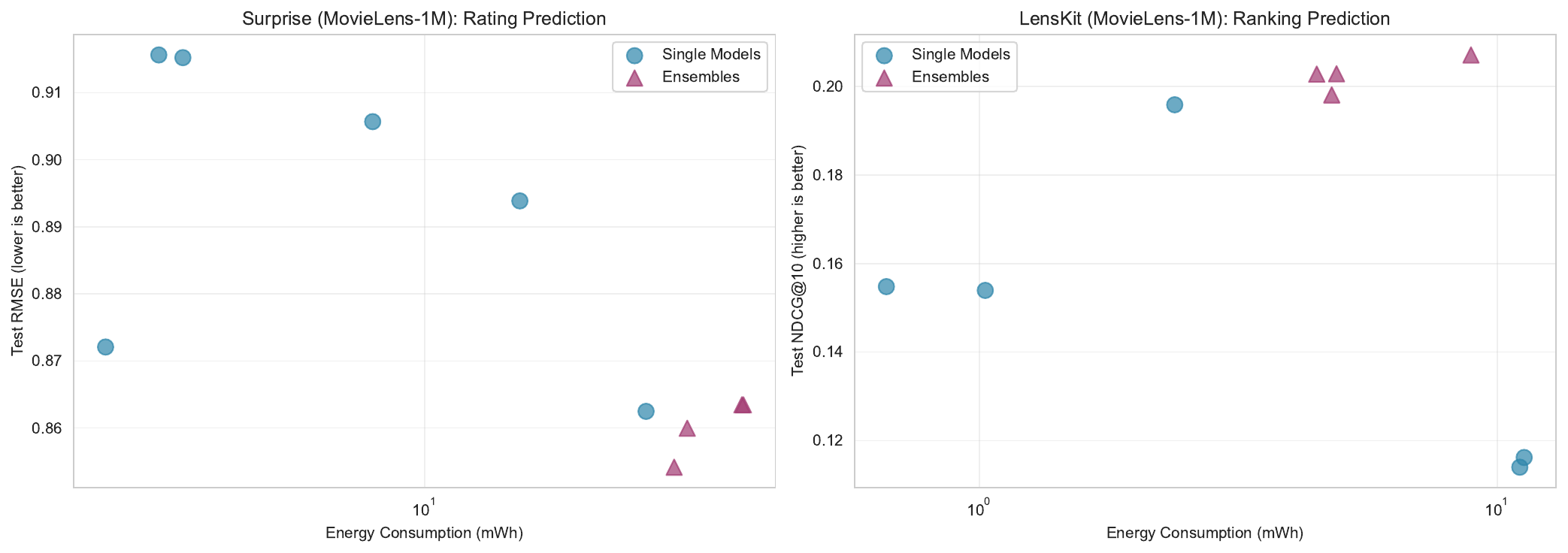}
    \caption{Comparison of accuracy-energy trade-offs between Surprise (rating prediction) and LensKit (ranking prediction) on MovieLens-1M. Both frameworks show ensemble methods consuming more energy while providing modest accuracy improvements. Note the logarithmic energy scale.}
    \label{fig:tradeoff_comparison}
\end{figure}

Figure~\ref{fig:tradeoff_comparison} provides a direct comparison of accuracy-energy trade-offs between frameworks on the MovieLens-1M dataset. Both frameworks show similar patterns: ensemble methods cluster in higher energy consumption regions (2-100 mWh) while providing marginal accuracy improvements over the best single models. Single models span a wider range of energy consumption (0.1-50 mWh) with varied accuracy performance.

\section{Carbon Footprint Analysis}

When examining individual datasets, the carbon footprint varies substantially. On MovieLens-100K, the difference between single models and ensembles is minimal (sub-milligram scale). On MovieLens-1M, ensembles produce 10-18 mg CO\textsubscript{2} compared to 0.4-10 mg CO\textsubscript{2} for single models. On the Anime dataset, where applicable, the Top Performers ensemble for Surprise produced 53.8 mg CO\textsubscript{2} compared to 2.6 mg CO\textsubscript{2} for the best single model (SVD), a 20-fold increase.

\section{Key Findings}

Across 93 successful experiments, several consistent patterns emerged:
\\
\\
\noindent \textbf{Accuracy improvements:} Ensemble methods achieved 0.3\% to 5.7\% accuracy improvements over the best single model, depending on dataset and framework. The largest improvements occurred on smaller datasets (MovieLens-100K: 5.7\% NDCG improvement) while larger datasets showed smaller gains (MovieLens-1M: 1.0\% RMSE improvement).
\\
\\
\noindent \textbf{Energy overhead:} Ensemble methods consumed 19\% to 2,549\% more energy than the best single model, with overhead correlating to dataset size and ensemble complexity. Simple ensembles (Average, Weighted) showed lower overhead than complex strategies (Stacking, Rank Fusion) on smaller datasets but comparable overhead on larger datasets.
\\
\\
\noindent\textbf{Model-specific efficiency:} Certain single models (SVD in Surprise, ALS in Lens\-Kit) demonstrated favorable accuracy-energy profiles, achieving competitive accuracy with low energy consumption, making them attractive alternatives to ensemble approaches.

\chapter{Discussion}

This chapter interprets the empirical findings presented in Chapter 5, examining their implications for sustainable recommender system design. We analyze the fundamental patterns underlying the observed trade-offs and evaluate scalability challenges.

\section{Non-Linear Accuracy-Energy Trade-Offs}

The most striking finding from our experiments is the highly non-linear relationship between accuracy improvements and energy consumption. Small gains in recommendation quality require disproportionately large increases in computational resources. The trade-off varies dramatically by dataset and framework: on MovieLens-100K, a 0.42\% RMSE improvement required 53\% more energy, while achieving 5.7\% NDCG improvement cost 103\% additional energy. This relationship challenges the implicit assumption in recommender systems research that accuracy improvements are always worth pursuing regardless of computational cost.

\subsection{Why Ensembles Are Energy-Intensive}

The energy overhead of ensemble methods is caused by three factors. First, training multiple base models multiplies computational costs, even efficient models like SVD become expensive when trained five or six times. Second, certain base models (SVD++, KNN-Baseline) are inherently expensive, and their inclusion in ensembles dominates overall energy consumption. Third, making predictions with ensembles requires running all base models and combining their outputs, which adds overhead to every recommendation request.
\\
\\
\noindent This explains why simple averaging strategies often consumed more energy than stacking: they invoke all base models including expensive ones, while stacking can learn to down-weight computationally expensive models that provide limited marginal benefit.

\subsection{Dataset Characteristics Matter}

The accuracy-energy trade-off varies substantially with dataset properties. Smaller datasets in the LensKit framework showed the largest accuracy improvements from ensembles (MovieLens-100K: 5.7\% NDCG improvement) but also high relative energy overheads (103\%). This happens because single models already capture most patterns in small datasets, ensembles only find small improvements by combining subtle differences, which requires additional computation for diminishing returns.
\\
\\
\noindent Larger, more complex datasets showed more favorable trade-offs. MovieLens-1M demonstrated that ensembles can achieve meaningful accuracy gains without proportionally scaling energy consumption when sufficient signal exists for complementary models to capture distinct patterns. However, the Anime dataset revealed hard scalability limits where memory constraints prevent ensemble execution entirely.

\subsection{Selective Combination as a Sustainable Strategy}

The Top-Performers ensemble consistently demonstrated superior energy efficiency by combining only 2-3 high-performing models instead of all available models. This selective approach reduces computational redundancy while preserving most ensemble benefits. The implication for sustainable ensemble design is clear: exhaustive combination is wasteful. Identifying and combining only complementary high-performers offers a more efficient path forward.

\section{Scalability}

The Anime dataset experiments exposed critical limitations that constrain ensemble applicability to production systems. Memory-intensive algorithms and ensemble aggregation both failed at this scale, revealing that ensemble techniques may not scale to industrial deployments without fundamental architectural changes.

\subsection{When Simple Models Win}

ModCloth results demonstrate that ensemble methods do not guarantee superiority. The simple Baseline model achieved competitive accuracy with negligible energy consumption, while KNN-Baseline severely overfitted. On sparse datasets with limited signal, adding more models introduces noise rather than capturing complementary patterns. This suggests researchers should always benchmark against simple baselines before investing in ensemble complexity.

\section{Carbon Footprint}

\subsection{The Serving Cost Problem}

Our measurements captured training energy, but ensemble overhead extends to recommendation time. Averaging and weighting strategies showed higher prediction latency because they must query multiple models per recommendation request. For systems serving millions of real-time recommendations, this serving cost dominates training overhead. Ensemble techniques work better for batch systems that pre-compute recommendations offline, where the training cost is spread across many cached predictions and latency is less critical.

\section{Practical Decision Framework}

When should practitioners choose ensembles over single models? The answer depends on several factors:

\noindent \textbf{Dataset scale and complexity:} Large datasets with rich interaction patterns offer better accuracy-energy trade-offs. Small or sparse datasets favor single optimized models or simple baselines.

\noindent \textbf{Accuracy requirements:} Critical domains (medical recommendations, financial advisory) may justify energy costs for improved accuracy. General-purpose entertainment recommendations may not.

\noindent \textbf{Organizational constraints:} Large technology companies with substantial computational budgets can afford ensemble overhead for marginal accuracy gains. In contrast, organizations with environmental commitments or limited resources should prioritize energy-efficient single models that deliver competitive performance at lower cost.

\subsection{Model Selection Guidelines}

Beyond ensemble decisions, framework-specific insights guide model selection. For explicit rating prediction, SVD offers exceptional efficiency, while achieving competitive accuracy with minimal energy consumption. For implicit feedback ranking, ALS demonstrates favorable profiles despite lower absolute accuracy than User-KNN. The sustainable choice often involves accepting modest accuracy reductions for substantial energy savings.

\chapter{Conclusion}

This thesis systematically evaluated the environmental impact of ensemble techniques in recommender systems, addressing a critical gap in sustainable machine learning research. Through controlled experiments across two frameworks and four benchmark datasets, we quantified the accuracy-energy trade-offs inherent in ensemble-based recommendation.

\section{Research Contributions}

Our work makes three primary contributions to the field of sustainable recommender systems:

\noindent \textbf{First}, we provide the first systematic measurement of energy consumption and carbon footprint for ensemble techniques in recommender systems. Previous research focused on accuracy metrics~\citep{ensemble_greedy_mehta2024, ensemble_graph_embedding_forouzandeh2020}, leaving researchers without data to evaluate sustainability trade-offs. Our experiments demonstrate that ensemble methods consume substantially more energy than single models, with modest accuracy improvements. This finding reveals a highly non-linear relationship where marginal accuracy gains require disproportionate energy investments.

\noindent \textbf{Second}, we demonstrate that ensemble strategy selection significantly impacts energy efficiency. The Top-Performers ensemble, which selectively combines only high-performing models, consistently exhibited superior energy efficiency compared to exhaustive averaging strategies. This insight provides actionable guidance: selective model combination offers a more sustainable path than indiscriminate combination of all available models.

\noindent \textbf{Third}, we identify critical scalability limitations that constrain ensemble applicability to large-scale systems. Memory-intensive algorithms and ensemble aggregation both failed on our largest dataset, highlighting fundamental challenges in deploying ensemble methods at industrial scale without distributed computing infrastructure.

\section{Implications for Practice}

Our findings have immediate practical implications for researchers and practitioners designing recommender systems:

\begin{itemize}
    \item \textbf{Report energy alongside accuracy:} The recommender systems community should adopt energy and carbon reporting as standard practice. Accuracy metrics alone provide an incomplete picture of algorithmic performance.
    
    \item \textbf{Consider single-model optimization:} For many applications, investing computational resources in optimizing single models through better hyperparameters, feature engineering, or data quality may lead to better accuracy-energy trade-offs than ensemble combination.
    
    \item \textbf{Apply ensembles selectively:} When ensemble techniques are justified, selective strategies offer the most sustainable approach. Combining carefully chosen models preserves ensemble benefits while minimizing computational redundancy.
    
    \item \textbf{Evaluate context-specific trade-offs:} The viability of ensemble techniques depends heavily on deployment context. Is accuracy critical? What are organizational energy constraints? Understanding these factors is essential for informed algorithm selection.
\end{itemize}

\section{Broader Impact}

Beyond recommender systems, our research contributes to the growing movement toward sustainable artificial intelligence~\citep{green_ai_schwartz2019}. The machine learning community increasingly recognizes that algorithmic progress cannot ignore environmental externalities. Our work provides a methodology for quantifying accuracy-energy trade-offs applicable to other domains where ensemble techniques are prevalent, like computer vision, natural language processing, and time series forecasting.
\\
\\
\noindent The urgency of climate action demands that we critically examine computational practices. With machine learning's carbon footprint growing rapidly~\citep{computing_carbon_gupta2021}, algorithmic design choices carry ethical weight. Our findings challenge the assumption that more computation always leads to better outcomes.

\section{Final Remarks}

Our research question asked: \textit{How do ensemble techniques in recommender systems influence the environmental impact compared to a single optimized model?} The answer is clear: Ensemble techniques substantially increase environmental impact through non-linear energy-accuracy relationships. Small accuracy gains require disproportionate energy investments.

\noindent However, context matters. Selective strategies applied to large, complex datasets in appropriate deployment contexts can achieve meaningful improvements with manageable overhead. The key insight is that ensemble techniques require careful evaluation of trade-offs rather than automatic deployment.

\noindent Sustainable recommender system design demands explicit consideration of environmental impact. Accuracy improvements must be weighed against their energy and carbon costs. Our research provides the empirical foundation for making these evaluations, demonstrating that environmentally responsible recommendation is both feasible and necessary. As the machine learning community confronts its role in climate change, measuring and minimizing the environmental impact of algorithmic choices becomes not just good practice, but an ethical imperative.

\chapter{Summary}

This thesis investigated the environmental impact of ensemble techniques in recommender systems compared to single optimized models. While machine learning's carbon footprint grows rapidly, ensemble methods have achieved substantial accuracy improvements (10-30\%) in recent research~\citep{ensemble_greedy_mehta2024, ensemble_graph_embedding_forouzandeh2020}, yet without measuring their energy consumption or carbon emissions.
\\
\\
\noindent We conducted systematic experiments across two frameworks: Surprise for explicit rating prediction (measured by RMSE) and LensKit for implicit feedback ranking (measured by NDCG@10). Four benchmark datasets of varying scales were evaluated: MovieLens-100K, MovieLens-1M, ModCloth, and Anime. Four ensemble strategies (Average, Weighted, Stacking/Rank Fusion, Top Performers) were compared against baseline and optimized single models. Energy consumption was measured using the EMERS tool with a smart plug, capturing total system power draw during training and prediction.
\\
\\
\noindent Results revealed a highly non-linear accuracy-energy relationship. Ensemble methods achieved modest accuracy improvements of 0.3-5.7\% over the best single models, while consuming 19-2,549\% more energy depending on dataset size and ensemble strategy. On MovieLens-1M, the Top Performers ensemble improved accuracy by 0.96\% (RMSE) while consuming 18.8\% more energy than the best single model. However, on smaller datasets like MovieLens-100K, ensembles achieved 5.7\% accuracy improvement (NDCG) with 103\% energy overhead. Scalability limitations emerged on the largest dataset (Anime), where memory constraints prevented ensemble execution entirely.
\\
\\
\noindent Three key contributions emerged: (1) first systematic measurement of energy and carbon footprint for ensemble recommender systems, (2) demonstration that selective ensemble strategies (Top Performers) offer superior energy efficiency compared to exhaustive averaging, and (3) identification of critical scalability limitations constraining ensemble applicability at industrial scale.
\\
\\
\noindent The findings have direct implications for sustainable algorithm selection: report energy alongside accuracy metrics, consider single-model optimization as an alternative to ensembles, apply ensemble techniques selectively rather than indiscriminately, and evaluate context-specific trade-offs based on deployment requirements. This research provides the empirical foundation for making informed decisions about ensemble deployment in recommender systems while minimizing environmental impact.

\chapter{Future Work and Limitations}

This chapter acknowledges the constraints of our research and identifies directions for future investigation to contextualize our findings and guide future work in sustainable recommender systems.

\section{Limitations}

\subsection{Scope and Generalizability}

Our experiments focused on four benchmark datasets (movies, fashion e-commerce) using traditional algorithms within Surprise and LensKit frameworks. This scope limits generalizability in several ways. First, other domains (music, news, social media) may exhibit different accuracy-energy relationships. Second, production systems operate at scales orders of magnitude larger than our largest dataset, where ensemble techniques may be even less viable. Third, deep learning models, graph-based methods, and more sophisticated ensemble strategies (boosting, cascade ensembles) remain unexplored.

\section{Future Work}

Future research should address these limitations through: (1) evaluating ensemble techniques across diverse domains and modern architectures like neural collaborative filtering and transformers; (2) developing adaptive ensembles that selectively invoke models based on prediction confidence; (3) investigating hardware optimization (GPUs, TPUs, quantization) and multi-objective optimization techniques; (4) collaborating with industry partners for production-scale validation; and (5) designing novel algorithms explicitly optimized for sustainability through learned model selection and resource sharing.

\section{Closing Remarks}

Despite these limitations, our research provides the first systematic evaluation of ensemble techniques' environmental impact in recommender systems. As the machine learning community increasingly prioritizes sustainability alongside accuracy, our work establishes a foundation and methodology demonstrating that rigorous measurement of accuracy-energy trade-offs is both feasible and necessary for sustainable algorithmic design.

\chapter{Acknowledgments} I would like to thank Prof. Dr. Joeran Beel, M.Sc. Lukas Wegmeth and B.Sc. Moritz Baumgart for their feedback and the helpful tips they provided during the preparation of this thesis and the associated short paper.
\\
\\
\noindent During the writing of this thesis, AI and large language model tools such as ChatGPT were used for translation, spell-checking, and proofreading. All content was still manually and carefully checked for correctness, and the ideas presented in this paper are original to me. My usage of such tools aligns with the usage guidelines by the Intelligent Systems Group (ISG) of the University of Siegen \cite{Beel2024a}.

\newpage
\renewcommand{\bibname}{References}
\addcontentsline{toc}{chapter}{References}
\bibliography{bibliography}{}

@misc{ensemble_greedy_mehta2024,
  title      = {Ensemble Boost: Greedy Selection for Superior Recommender Systems},
  url        = {http://arxiv.org/abs/2407.05221},
  doi        = {10.48550/arXiv.2407.05221},
  shorttitle = {Ensemble Boost},
  number     = {arXiv:2407.05221},
  publisher  = {arXiv},
  author     = {Mehta, Zainil and Vente, Tobias},
  date       = {2024-07-07},
  eprinttype = {arxiv},
  eprint     = {2407.05221}
}

@misc{ensemble_graph_embedding_forouzandeh2020,
  title      = {Presentation of a Recommender System with Ensemble Learning and Graph Embedding: A Case on MovieLens},
  url        = {http://arxiv.org/abs/2008.01192},
  doi        = {10.48550/arXiv.2008.01192},
  number     = {arXiv:2008.01192},
  publisher  = {arXiv},
  author     = {Forouzandeh, Saman and Rostami, Mehrdad and Berahmand, Kamal},
  date       = {2020-07-15},
  eprinttype = {arxiv},
  eprint     = {2008.01192}
}

@misc{ensemble_model_selection_nijkamp2025,
  title      = {Green AI in Action: Strategic Model Selection for Ensembles in Production},
  url        = {http://arxiv.org/abs/2405.17451},
  doi        = {10.48550/arXiv.2405.17451},
  number     = {arXiv:2405.17451},
  publisher  = {arXiv},
  author     = {Nijkamp, Nienke and Sallou, June and van der Heijden, Niels and Cruz, Luís},
  date       = {2025-06-30},
  eprinttype = {arxiv},
  eprint     = {2405.17451}
}

@misc{ensemble_accuracy_energy_omar2024,
  title      = {The More the Merrier? Navigating Accuracy vs. Energy Efficiency Design Trade-Offs in Ensemble Learning Systems},
  url        = {http://arxiv.org/abs/2407.02914},
  doi        = {10.48550/arXiv.2407.02914},
  number     = {arXiv:2407.02914},
  publisher  = {arXiv},
  author     = {Omar, Rafiullah and Bogner, Justus and Muccini, Henry and Lago, Patricia and Martínez-Fernández, Silverio and Franch, Xavier},
  date       = {2024-07-03},
  eprinttype = {arxiv},
  eprint     = {2407.02914}
}

@inproceedings{recsys_carbon_footprint_vente2024,
  location  = {New York, NY, USA},
  title     = {From Clicks to Carbon: The Environmental Toll of Recommender Systems},
  isbn      = {979-8-4007-0505-2},
  url       = {https://doi.org/10.1145/3640457.3688074},
  doi       = {10.1145/3640457.3688074},
  series    = {RecSys '24},
  pages     = {580--590},
  booktitle = {Proceedings of the 18th ACM Conference on Recommender Systems},
  publisher = {Association for Computing Machinery},
  author    = {Vente, Tobias and Wegmeth, Lukas and Said, Alan and Beel, Joeran},
  date      = {2024}
}

@inproceedings{recsys_sustainability_tradeoff_spillo2023,
  location  = {Singapore},
  title     = {Towards Sustainability-aware Recommender Systems: Analyzing the Trade-off Between Algorithms Performance and Carbon Footprint},
  isbn      = {979-8-4007-0241-9},
  url       = {https://dl.acm.org/doi/10.1145/3604915.3608840},
  doi       = {10.1145/3604915.3608840},
  pages     = {856--862},
  booktitle = {Proceedings of the 17th ACM Conference on Recommender Systems},
  publisher = {ACM},
  author    = {Spillo, Giuseppe and De Filippo, Allegra and Musto, Cataldo and Milano, Michela and Semeraro, Giovanni},
  date      = {2023-09-14}
}

@misc{dataset_size_energy_arabzadeh2024,
  title      = {Green Recommender Systems: Optimizing Dataset Size for Energy-Efficient Algorithm Performance},
  url        = {http://arxiv.org/abs/2410.09359},
  doi        = {10.48550/arXiv.2410.09359},
  number     = {arXiv:2410.09359},
  publisher  = {arXiv},
  author     = {Arabzadeh, Ardalan and Vente, Tobias and Beel, Joeran},
  date       = {2024-11-05},
  eprinttype = {arxiv},
  eprint     = {2410.09359}
}

@misc{carbon_aware_recsys_kalisvaart2025,
  title      = {Towards Carbon Footprint-Aware Recommender Systems for Greener Item Recommendation},
  url        = {http://arxiv.org/abs/2503.17201},
  doi        = {10.48550/arXiv.2503.17201},
  number     = {arXiv:2503.17201},
  publisher  = {arXiv},
  author     = {Kalisvaart, Raoul and Mansoury, Masoud and Hanjalic, Alan and Isufi, Elvin},
  date       = {2025-03-21},
  eprinttype = {arxiv},
  eprint     = {2503.17201}
}

@misc{gnn_sustainable_purificato2024,
  title      = {Eco-Aware Graph Neural Networks for Sustainable Recommendations},
  url        = {http://arxiv.org/abs/2410.09514},
  doi        = {10.48550/arXiv.2410.09514},
  number     = {arXiv:2410.09514},
  publisher  = {arXiv},
  author     = {Purificato, Antonio and Silvestri, Fabrizio},
  date       = {2024-10-12},
  eprinttype = {arxiv},
  eprint     = {2410.09514}
}

@article{recsys_sustainability_overview_felfernig2023,
  title        = {Recommender systems for sustainability: overview and research issues},
  volume       = {6},
  issn         = {2624-909X},
  url          = {https://www.frontiersin.org/articles/10.3389/fdata.2023.1284511/full},
  doi          = {10.3389/fdata.2023.1284511},
  pages        = {1284511},
  journaltitle = {Frontiers in Big Data},
  author       = {Felfernig, Alexander and Wundara, Manfred and Tran, Thi Ngoc Trang and Polat-Erdeniz, Seda and Lubos, Sebastian and El Mansi, Merfat and Garber, Damian and Le, Viet-Man},
  date         = {2023-10-30}
}

@article{random_seeds_wegmeth2024,
  title  = {The Effect of Random Seeds for Data Splitting on Recommendation Accuracy},
  author = {Wegmeth, Lukas and Vente, Tobias and Purucker, Lennart and Beel, Joeran},
  year   = {2024}
}

@misc{recsys_comprehensive_review_raza2025,
  title  = {A Comprehensive Review of Recommender Systems: Transitioning from Theory to Practice},
  url    = {https://arxiv.org/abs/2407.13699},
  author = {Raza, Shaina and Rahman, Mizanur and Kamawal, Safiullah and Toroghi, Armin and Raval, Ananya and Navah, Farshad and Kazemeini, Amirmohammad},
  date   = {2025},
  note   = {eprint: 2407.13699}
}

@misc{green_ai_schwartz2019,
  title      = {Green AI},
  url        = {http://arxiv.org/abs/1907.10597},
  doi        = {10.48550/arXiv.1907.10597},
  number     = {arXiv:1907.10597},
  publisher  = {arXiv},
  author     = {Schwartz, Roy and Dodge, Jesse and Smith, Noah A. and Etzioni, Oren},
  date       = {2019-08-13},
  eprinttype = {arxiv},
  eprint     = {1907.10597}
}

@misc{ml_energy_reporting_henderson2020,
  title      = {Towards the Systematic Reporting of the Energy and Carbon Footprints of Machine Learning},
  url        = {http://arxiv.org/abs/2002.05651},
  author     = {Henderson, Peter and Hu, Jieru and Romoff, Joshua and Brunskill, Emma and Jurafsky, Dan and Pineau, Joelle},
  date       = {2020},
  eprinttype = {arxiv},
  eprint     = {2002.05651}
}

@inproceedings{computing_carbon_gupta2021,
  title     = {Chasing Carbon: The Elusive Environmental Footprint of Computing},
  url       = {https://ieeexplore.ieee.org/document/9407142/},
  doi       = {10.1109/HPCA51647.2021.00076},
  pages     = {854--867},
  booktitle = {2021 IEEE International Symposium on High-Performance Computer Architecture (HPCA)},
  author    = {Gupta, Udit and Kim, Young Geun and Lee, Sylvia and Tse, Jordan and Lee, Hsien-Hsin S. and Wei, Gu-Yeon and Brooks, David and Wu, Carole-Jean},
  date      = {2021-02}
}

@article{climate_consensus_lynas2021,
  title        = {Greater than 99\% consensus on human caused climate change in the peer-reviewed scientific literature},
  volume       = {16},
  issn         = {1748-9326},
  url          = {https://iopscience.iop.org/article/10.1088/1748-9326/ac2966},
  doi          = {10.1088/1748-9326/ac2966},
  pages        = {114005},
  number       = {11},
  journaltitle = {Environmental Research Letters},
  author       = {Lynas, Mark and Houlton, Benjamin Z and Perry, Simon},
  date         = {2021-10-19}
}

@misc{emers_tool_wegmeth2024,
  title      = {EMERS: Energy Meter for Recommender Systems},
  url        = {http://arxiv.org/abs/2409.15060},
  doi        = {10.48550/arXiv.2409.15060},
  number     = {arXiv:2409.15060},
  publisher  = {arXiv},
  author     = {Wegmeth, Lukas and Vente, Tobias and Said, Alan and Beel, Joeran},
  date       = {2024-09-23},
  eprinttype = {arxiv},
  eprint     = {2409.15060}
}

@software{emers_github,
  title     = {ISG-Siegen/emers},
  rights    = {MIT},
  url       = {https://github.com/ISG-Siegen/emers},
  publisher = {Intelligent Systems Group, University of Siegen},
  date      = {2025-05-13}
}

@inproceedings{power_meters_comparison_jay2023,
  title     = {An experimental comparison of software-based power meters: focus on CPU and GPU},
  url       = {https://ieeexplore.ieee.org/document/10171575/},
  doi       = {10.1109/CCGrid57682.2023.00020},
  pages     = {106--118},
  booktitle = {2023 IEEE/ACM 23rd International Symposium on Cluster, Cloud and Internet Computing (CCGrid)},
  author    = {Jay, Mathilde and Ostapenco, Vladimir and Lefevre, Laurent and Trystram, Denis and Orgerie, Anne-Cécile and Fichel, Benjamin},
  date      = {2023-05}
}

@article{ndcg_metric_jarvelin2002,
  title        = {Cumulated gain-based evaluation of IR techniques},
  volume       = {20},
  issn         = {1046-8188},
  url          = {https://dl.acm.org/doi/10.1145/582415.582418},
  doi          = {10.1145/582415.582418},
  pages        = {422--446},
  number       = {4},
  journaltitle = {ACM Trans. Inf. Syst.},
  author       = {Järvelin, Kalervo and Kekäläinen, Jaana},
  date         = {2002-10-01}
}

@misc{ndcg_theory_wang2013,
  title      = {A Theoretical Analysis of NDCG Type Ranking Measures},
  url        = {http://arxiv.org/abs/1304.6480},
  doi        = {10.48550/arXiv.1304.6480},
  number     = {arXiv:1304.6480},
  publisher  = {arXiv},
  author     = {Wang, Yining and Wang, Liwei and Li, Yuanzhi and He, Di and Liu, Tie-Yan and Chen, Wei},
  date       = {2013-04-24},
  eprinttype = {arxiv},
  eprint     = {1304.6480}
}

@inproceedings{recbole_framework_zhao2021,
  title     = {RecBole: Towards a Unified, Comprehensive and Efficient Framework for Recommendation Algorithms},
  pages     = {4653--4664},
  booktitle = {CIKM},
  publisher = {ACM},
  author    = {Zhao, Wayne Xin and Mu, Shanlei and Hou, Yupeng and Lin, Zihan and Chen, Yushuo and Pan, Xingyu and Li, Kaiyuan and Lu, Yujie and Wang, Hui and Tian, Changxin and Min, Yingqian and Feng, Zhichao and Fan, Xinyan and Chen, Xu and Wang, Pengfei and Ji, Wendi and Li, Yaliang and Wang, Xiaoling and Wen, Ji-Rong},
  date      = {2021}
}

@misc{recbole_datasets,
  title  = {RecBole Datasets},
  url    = {https://drive.google.com/drive/folders/1so0lckI6N6_niVEYaBu-LIcpOdZf99kj},
  author = {{RecBole}},
  date   = {2025-07-20}
}

@online{lenskit_documentation,
  title = {LensKit Documentation},
  url   = {https://lenskit.org/stable/index.html},
  date  = {2025-08-01}
}

@article{surprise_joss_hug2020,
  title   = {Surprise: A Python library for recommender systems},
  volume  = {5},
  url     = {https://joss.theoj.org/papers/10.21105/joss.02174},
  doi     = {10.21105/joss.02174},
  number  = {52},
  journal = {Journal of Open Source Software},
  author  = {Hug, Nicolas},
  year    = {2020},
  pages   = {2174}
}

@article{movielens_dataset_harper2015,
  author    = {Harper, F. Maxwell and Konstan, Joseph A.},
  title     = {The MovieLens Datasets: History and Context},
  year      = {2015},
  volume    = {5},
  number    = {4},
  issn      = {2160-6455},
  url       = {https://doi.org/10.1145/2827872},
  doi       = {10.1145/2827872},
  journal   = {ACM Trans. Interact. Intell. Syst.},
  articleno = {19},
  numpages  = {19}
}

@article{crossfold_bengio_2004,
  title        = {No Unbiased Estimator of the Variance of K-Fold Cross-Validation},
  volume       = {5},
  issn         = {{ISSN} 1533-7928},
  url          = {https://jmlr.csail.mit.edu/papers/v5/grandvalet04a.html},
  pages        = {1089--1105},
  issue        = {Sep},
  journaltitle = {Journal of Machine Learning Research},
  author       = {Bengio, Yoshua and Grandvalet, Yves},
  urldate      = {2025-10-25},
  date         = {2004}
}

@online{thesis_github_repo,
  title      = {ISG-Siegen/BA\_Jannik\_Nitschke},
  url        = {https://github.com/ISG-Siegen/BA_Jannik_Nitschke},
  titleaddon = {GitHub},
  date       = {2025}
}

@inproceedings{Beel2024a,
  author    = {Joeran Beel},
  booktitle = {Intelligent Systems Group, Blog},
  title     = {Our use of AI-tools for writing research papers},
  year      = {2024},
  url       = {https://isg.beel.org/blog/2024/08/19/our-use-of-ai-tools-for-writing-research-papers/}
}
\bibliographystyle{plainnat}

\end{document}